\pdfoutput=1
\documentclass[conference]{IEEEtran}
\IEEEoverridecommandlockouts

\PassOptionsToPackage{numbers,sort&compress}{natbib}
\usepackage{natbib}

\usepackage[utf8]{inputenc}
\usepackage[T1]{fontenc}
\usepackage{amsmath,amssymb,amsfonts}
\usepackage{graphicx}
\usepackage{booktabs}
\usepackage{multirow}
\usepackage{xcolor}
\usepackage{url}
\usepackage{xspace}
\usepackage{microtype}
\usepackage[hypcap=true]{caption}
\usepackage{hyperref}

\usepackage{algorithm}
\usepackage{algpseudocode}
\usepackage{siunitx}

\usepackage{tikz}
\usetikzlibrary{positioning, shapes.geometric, arrows.meta, fit}
\usepackage{pgfplots}
\pgfplotsset{compat=1.18}
\usepgfplotslibrary{groupplots}
\usetikzlibrary{calc}
\usetikzlibrary{patterns}

\colorlet{cSys}{blue!70!black}
\colorlet{cOpus}{red!70!black}
\colorlet{cGpt}{teal!70!black}
\colorlet{cKimi}{violet!70!black}
\colorlet{cFlash}{orange!70!black}
\colorlet{cBase}{gray}

\pgfplotsset{
  paperaxis/.style={
    width=0.80\linewidth,
    height=5.2cm,
    tick label style={font=\footnotesize},
    label style={font=\footnotesize},
    title style={font=\small},
    legend style={font=\footnotesize, draw=gray!50, fill=white,
                  fill opacity=0.9, text opacity=1, rounded corners=1pt},
    every axis plot/.append style={line width=0.7pt},
    grid style={gray!25, line width=0.3pt},
    axis line style={gray!60},
    tick style={gray!60},
  },
  papermark/.style={mark size=2.6pt},
}

\usepackage{placeins}
\usepackage{float}

\usepackage{balance}

\usepackage{titlesec}
\titleformat{\section}
  {\normalfont\large\bfseries}{\thesection}{0.8em}{}
\titleformat{\subsection}
  {\normalfont\normalsize\bfseries}{\thesubsection}{0.8em}{}
\titleformat{\subsubsection}
  {\normalfont\normalsize\bfseries\itshape}{\thesubsubsection}{0.8em}{}
\renewcommand{\thesection}{\arabic{section}}
\renewcommand{\thesubsection}{\arabic{section}.\arabic{subsection}}
\renewcommand{\thesubsubsection}{\arabic{section}.\arabic{subsection}.\arabic{subsubsection}}

\newcommand{\sysname}{SuperScout\xspace}
\newcommand{\modelname}{SuperScout-7B\xspace}

\newcommand{\fxgpt}{GPT-5.2\xspace}
\newcommand{\fxopus}{Claude Opus 4.6\xspace}
\newcommand{\fxflash}{Gemini 3 Flash\xspace}
\newcommand{\fxkimi}{Kimi K2.5\xspace}
\newcommand{\fxgptS}{GPT-5.2\xspace}
\newcommand{\fxopusS}{Opus 4.6\xspace}
\newcommand{\fxkimiS}{Kimi K2.5\xspace}

\newcommand{\code}[1]{\texttt{#1}}
\newcommand{\dataset}[1]{\textsc{#1}}
\title{Scrouting: Cost-Aware Routing of Coding Agents by Scouting the
Repository First%
\thanks{Model and data: \url{https://huggingface.co/SuperAGI/SuperScout-7B}
(weights), \url{https://huggingface.co/datasets/SuperAGI/superscout-sft-search}
(training corpus),
\url{https://huggingface.co/datasets/SuperAGI/superscout-router-features}
(router features). Code and data:
\url{https://github.com/TransformerOptimus/superscout}.}}

\author{%
\IEEEauthorblockN{\textbf{Mukunda NS}}
\IEEEauthorblockA{SuperAGI Research}
\and
\IEEEauthorblockN{\textbf{Adithyan Krishnan}}
\IEEEauthorblockA{SuperAGI Research}
\and
\IEEEauthorblockN{\textbf{Ishaan Bhola}}
\IEEEauthorblockA{SuperAGI Research}
\and
\IEEEauthorblockN{\textbf{Dhruv Jaglan}}
\IEEEauthorblockA{Rilo - IIT Bombay}
}

\begin{document}

% Activates the IEEEtranBSTCTL entry in references.bib: IEEE house rule of
% "et al." after six authors (first author shown). Renders nothing.
\bstctlcite{BSTcontrol}

\maketitle

% Abstract.

\begin{abstract}
Frontier language models can resolve repository-level software issues, but each
attempt is expensive, and existing routers select a model from the issue text
alone.  We present \sysname{}, which routes after scouting the repository: a 7B searcher,
\modelname{}, first explores the repository and produces a structured handoff
whose reproduction claims are sandbox-verified, with false claims stripped
before delivery.  The searcher's hidden states, together with the task text, then feed a
r\'esum\'e-based router that dispatches the task to one of four frontier
fixers.  Adding a new
fixer requires no retraining.  On the full Python slice of SWE-bench Pro (266
tasks) under the benchmark's official capped budget tier, \sysname{} matches
the best single model's solve rate (159 of 266 for \sysname{}, 158 for the
best model) at about a fifth of the
total cost per solve, and the reported configuration sits above the random
traffic-splitting baseline.  A no-router ablation, always the cheapest fixer
with the handoff, ties the routed system on this benchmark, so the handoff
rather than the routing decision carries the result.  A paired calibration
study points to the mechanism: the handoff appears to redistribute rather than
add solving ability, lifting the three cheaper fixers while slightly hurting
the strongest, though at $N{=}99$ the per-fixer effects are directional only;
the searcher's hidden states improve cost routing on the calibration labels
while the handoff's own text does not.  The searcher's compute adds less than half a cent of GPU time per
task.
\end{abstract}

% Section 1, Introduction.
\section{Introduction}
\label{sec:intro}

Frontier-model agents now resolve real repository issues, from locating
bugs to generating patches~\citep{swebench}, but each solve carries a
real per-task price.  Strong open-weights models cost a fraction yet
close fewer tasks.  Any team deploying an issue-solving system therefore
faces a standing choice between the expensive best model and cheaper
alternatives.  Existing LLM routers~\citep{routellm,frugalgpt} make
that choice from the task text alone, before anything has engaged with
the actual repository.

Replaying learned routers over the public per-task results of three SWE
benchmarks reveals that solve sets are largely nested: models that solve
more tasks nearly always subsume the solve sets of weaker ones, and no
router reliably improves on always selecting the strongest
model~(\S\ref{sec:problem}).  Routing for accuracy therefore offers
little headroom.  Routing for \emph{cost}, by contrast, only requires
predicting when a cheaper model will suffice.

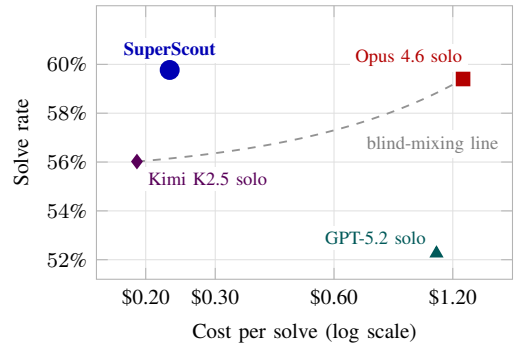
\begin{figure}[!tbp]
\centering
% ===========================================================================
% fig-cost-solve.tex — cost-per-solve vs solve-rate scatter, with the
% kimi<->opus blind-mixing line.
% BODY ONLY (tikzpicture). Wrapper/caption/label live in the section file.
% Requires \input{figures/fig-style} in the preamble.
% Provenance: data/receipts/outcomes_per_task.json and
%             data/receipts/routing_scenarios.json.
%
% Design notes:
%   - The router point and the no-router ablation differ by $0.003 per solve,
%     under 1.5pt on this log axis, so they share ONE composite glyph (open
%     cKimi diamond for the ablation, smaller filled cSys disc drawn last
%     inside it) with one colour-keyed callout and one leader; two markers
%     would misstate the resolution of the figure.
%   - The mixing reference is convex in LINEAR cost, so it is sampled as the
%     true mixture and renders as a curve on a log-x axis.
% ===========================================================================
\begin{tikzpicture}
\begin{axis}[
  paperaxis,
  xmode=log,
  log basis x=10,
  xmin=0.150, xmax=1.70,
  ymin=51.2, ymax=62.4,
  xtick={0.2,0.3,0.6,1.2},
  xticklabels={\$0.20,\$0.30,\$0.60,\$1.20},
  log ticks with fixed point,
  xminorticks=false,
  ytick={52,54,56,58,60},
  yticklabel={\pgfmathprintnumber{\tick}\%},
  xlabel={Cost per solve (log scale)},
  ylabel={Solve rate},
  ymajorgrids=true,
  xmajorgrids=true,
]

% --- blind-mixing locus: kimi -> opus (reference, deliberately subdued) --
% Solve rate is affine in LINEAR cost, so on a log-x axis this is a CURVE,
% not a straight segment; drawing it straight would overstate the baseline.
% Sampled densely; the endpoints are exactly the two solo points.
\addplot[dashed, line width=0.7pt, color=cBase!85, mark=none, forget plot,
         domain=0.190:1.274, samples=120, smooth]
  {56.02 + (x-0.190)*(59.40-56.02)/(1.274-0.190)};

% --- solo anchors (context) ---------------------------------------------
\addplot[only marks, mark size=2.4pt, color=cOpus, mark=square*, forget plot]
  coordinates {(1.274,59.40)};
\addplot[only marks, mark size=2.4pt, color=cKimi, mark=diamond*, forget plot]
  coordinates {(0.190,56.02)};
\addplot[only marks, mark size=2.4pt, color=cGpt, mark=triangle*, forget plot]
  coordinates {(1.091,52.26)};

% --- \sysname (protagonist): larger mark, drawn last, on top ------------
% The no-router ablation is deliberately NOT plotted here; it lives in the
% main results table. The figure carries the headline only.
\addplot[only marks, mark size=3.3pt, color=cSys, mark=*,
         mark options={fill=cSys, draw=cSys, line width=0.9pt}, forget plot]
  coordinates {(0.230,59.77)};
\node[font=\scriptsize, color=cSys, anchor=south, inner sep=1.5pt, fill=white]
  at (axis cs:0.230,60.05) {\textbf{\sysname}};

% --- inline labels ------------------------------------------------------
% Placement rule: each label is anchored so its box lies in a quadrant no
% marker, no grid tick and no segment of the dashed line passes through.
\node[font=\scriptsize, color=cOpus, anchor=south east, inner sep=2pt, fill=white]
  at (axis cs:1.310,59.70) {\fxopusS solo};
\node[font=\scriptsize, color=cKimi, anchor=north west, inner sep=2pt, fill=white]
  at (axis cs:0.196,55.80) {\fxkimiS solo};
\node[font=\scriptsize, color=cGpt, anchor=south east, inner sep=2pt, fill=white]
  at (axis cs:1.060,52.40) {\fxgptS solo};
\node[font=\scriptsize, color=cBase, anchor=north east, inner sep=2pt, fill=white]
  at (axis cs:1.620,57.30) {blind-mixing line};

\end{axis}
\end{tikzpicture}
\caption{\textbf{Cost per solve versus solve rate on SWE-bench Pro
(Python-266).}  The dashed curve is the blind-mixing line (random
cost-blind mix of \fxkimi and \fxopus).  \sysname matches \fxopus at
about a fifth of its cost per solve, well above the line.  The system
point is all-in; solo points are fixer API only.  The $x$-axis is
logarithmic.}
\label{fig:cost-solve}
\end{figure}

\sysname routes \emph{after} scouting, a pattern we call
\emph{scrouting}.  A 7B searcher,
\modelname, first explores the repository and produces a structured
handoff: a list of implicated files, diagnostic notes, and a candidate
reproduction test.  These claims pass through a sandbox verification
step, and claims that do not check out are stripped before a downstream
model ever sees them.
A r\'esum\'e router then selects among four frontier fixers using the
task text together with the searcher's own hidden states; adding a new
fixer requires no retraining.

Figure~\ref{fig:cost-solve} compares \sysname to solo frontier models on
SWE-bench Pro's full Python slice~\citep{swebenchpro} (266 tasks),
matching the benchmark's official capped budget tier exactly.  \sysname
matches the best single frontier model's solve rate, resolving 159 of
266 tasks versus 158, at \$0.230 total cost per solve compared to
\$1.274: about a fifth.  The configuration sits above the blind-mixing
line (the accuracy/cost segment any random split of traffic between a
cheap and a strong model would achieve).  The searcher's contribution to
system cost is negligible: \modelname's entire GPU bill for the
evaluation was \$1.13.

A paired calibration study on 100 fresh tasks points to why this cost
reduction holds: the handoff appears to redistribute rather than add
solving ability, lifting the three cheaper fixers while slightly hurting
the strongest, and the
searcher's hidden states improve cost routing where the handoff's own
text does not~(\S\ref{sec:calibration}).

We make five contributions:
\begin{itemize}
  \item An engage-then-route architecture with a trained searcher whose
    verified handoff is consumed by the chosen fixer.
  \item A zero-cost replay audit of published per-task outcomes on three
    SWE benchmarks showing that solve sets are largely nested and that no
    learned router reliably beats always calling the strongest model,
    which motivates routing for cost rather than accuracy.
  \item Routing features drawn from the searcher's hidden states, with a
    r\'esum\'e-based $N$-way pool where adding a new fixer requires no
    retraining.
  \item A matched-protocol evaluation on SWE-bench Pro's Python census
    showing frontier-matching accuracy at about a fifth of the total
    cost per solve.
  \item A paired calibration study measuring the handoff's
    redistribution pattern and the router's feature design space, plus a
    verification gate that strips the searcher's false reproduction
    claims.
\end{itemize}

% Section 2, Related Work.

\section{Related Work}
\label{sec:related}

SWE-agent~\citep{sweagent} introduced the agent-computer-interface paradigm
for autonomous issue resolution in real repositories.
Agentless~\citep{agentless} showed that a fixed pipeline, with no agent
autonomy at all, can achieve competitive resolve rates.
SWE-smith~\citep{swesmith} addresses data scarcity by synthesizing
large-scale training corpora for such agents, while
SWE-Gym~\citep{swegym} and R2E-Gym~\citep{r2egym} supply executable
training environments built from real repositories.
The community evaluates these systems on
SWE-bench~\citep{swebench} and its harder successor
SWE-bench Pro~\citep{swebenchpro}.

A parallel line of work trains small models specifically for code
localization.
SWE-Fixer~\citep{swefixer} trains a 7B retriever whose output feeds one
fixed larger editor.
LocAgent~\citep{locagent} and SweRank~\citep{swerank} similarly train
compact models to identify fault locations within a repository.
These systems produce exactly the kind of evidence a router could consume,
yet none of them routes: the localizer's output terminates in a single,
predetermined consumer.

LLM routing has been studied along two axes.
Cost-quality routers such as RouteLLM~\citep{routellm},
FrugalGPT~\citep{frugalgpt}, and Hybrid LLM~\citep{hybridllm} learn to
dispatch queries to cheaper or stronger models based on predicted difficulty.
Profile-based selectors take a complementary approach, building per-model
signatures from benchmark outcomes or learned
embeddings and matching incoming queries against
them~\citep{shnitzer,embedllm,iclrouter}.
Both families share a structural property: they decide from the task prompt,
and optionally from model profiles, before any system has engaged with the
concrete problem instance.

Several concurrent efforts address routing in code-generation settings.
SWE-Router~\citep{swerouter} probes each issue with a weak model and then
escalates between exactly two models; on escalation the strong model restarts
from scratch, so the probe's work is not consumed.
Its theoretical analysis of trajectory-conditioned routing nonetheless
supports the design direction we pursue.
CodeRescue~\citep{coderescue} routes among recovery actions (reflect,
replan, or escalate) inside a single agent's trajectory rather than among
fixer models.
TRACE-Router~\citep{tracerouter} formulates model selection as a contextual
bandit over an N-way pool, though it is evaluated outside software
engineering.
Self-play SWE-RL~\citep{selfplayswerl} trains solvers via self-play and
evaluates on SWE-bench Pro but does not route.
As Table~\ref{tab:capability-matrix} summarizes, \sysname is distinct in
combining a trained searcher, a handoff the fixer actually consumes (after
verification), an N-way fixer pool where adding a new fixer requires no
retraining, and routing features drawn from the searcher's hidden states.

\begin{table}[!tbp]
\centering
\caption{\textbf{Capability comparison of \sysname with related routing and
localization systems.}
Columns: \emph{Searcher}, a model trained to scout the repository before
any fix is attempted;
\emph{Handoff}, the searcher's work product is consumed by the downstream
fixer rather than discarded;
\emph{$N$-way}, selection over a pool of more than two fixers;
\emph{Onboard}, adding a new fixer requires no retraining of any learned
component;
\emph{States}, routing features include the searcher's internal hidden
states.
Concurrent code-routing work escalates between a fixed pair and discards
the cheap model's evidence on escalation;
profile-based routers select from the task prompt alone, before any
exploration;
trained small localizers produce evidence a router could use but perform no
routing.
$\checkmark^{\dagger}$: SWE-Fixer's retriever feeds one fixed larger editor,
while LocAgent and SweRank stop at localization.
Rows: SWE-Router~\citep{swerouter}, CodeRescue~\citep{coderescue},
TRACE-Router~\citep{tracerouter},
profile routers~\citep{embedllm,iclrouter,shnitzer},
small localizers~\citep{swefixer,locagent,swerank}.}
\label{tab:capability-matrix}
\begin{tabular}{@{}lccccc@{}}
\toprule
System & Searcher & Handoff & $N$-way & Onboard & States \\
\midrule
SWE-Router      & $\times$     & $\times$ & $\times$     & $\times$     & $\times$ \\
CodeRescue      & $\times$     & $\times$ & $\times$     & $\times$     & $\times$ \\
TRACE-Router    & $\times$     & $\times$ & $\checkmark$ & $\times$     & $\times$ \\
Profile routers & $\times$     & $\times$ & $\checkmark$ & $\checkmark$ & $\times$ \\
Small localizers & $\checkmark$ & $\checkmark^{\dagger}$ & $\times$ & $\times$ & $\times$ \\
\midrule
\textbf{\sysname} & $\checkmark$ & $\checkmark$ & $\checkmark$ & $\checkmark$ & $\checkmark$ \\
\bottomrule
\end{tabular}

\end{table}
\FloatBarrier

% Section 3, Problem Setup and Metrics.
\section{Problem Setup and Metrics}
\label{sec:problem}

\subsection{Task and Cost Model}
\label{sec:problem:task}

We study repo-level issue resolution: given the text of an issue and a
snapshot of the repository, the system must produce a source-code patch
whose correctness is judged by held-out tests~\citep{swebench,swebenchpro}.
Frontier language models already solve these tasks at high rates, but their
per-task cost is substantial.  Strong open-weights models are far cheaper
yet less reliable.  In a deployment that selects from a pool of candidate
\emph{fixer} models, the natural question is whether one can retain the
strongest model's effectiveness while spending less.

To answer that question we adopt a single yardstick, defined before any
experiment: \textbf{total cost per solve at a matched solve rate}.
``Total'' means every system component's spend, including searcher GPU
time, verification sandboxes, and fixer API calls, not just the final
model invocation.  A system claims a cost improvement only if its solve
rate matches that of the best individual model.

This metric exposes a trivial lower bound.  Any point on the line segment
between a cheap model's (cost, accuracy) and a strong model's (cost,
accuracy) is achievable by randomly splitting traffic between the two, a
strategy we call \emph{blind mixing}.  A routing system is interesting
only if it operates above that segment.  We describe the operational
details of the blind-mixing line in~\S\ref{sec:setup}.

\subsection{Why Accuracy-Only Routing Is Insufficient}
\label{sec:problem:audit}

Before building \sysname we replayed a set of learned routers over the
publicly available per-task results of three SWE-coding benchmarks:
SWE-bench Verified~\citep{swebenchverified}, SWE-bench
Multilingual~\citep{swebenchmultilingual}, and SWE-bench
Pro~\citep{swebenchpro}.  The replay incurs zero model cost, and the
routers span four families: embedding retrieval ($k$-NN), trained heads
(logistic and MLP), clustering, and a language-rule baseline.  Two
findings shaped the design that followed.

\begin{figure}[!tbp]
\centering
\resizebox{\linewidth}{!}{%
\begin{tikzpicture}[
  x=1cm, y=1cm,
  lbl/.style={font=\footnotesize, text=gray!25!black, align=center,
              inner sep=0pt},
  micro/.style={font=\scriptsize, text=gray!45!black, align=center,
                inner sep=0pt},
  kick/.style={font=\scriptsize\scshape, text=gray!45!black, inner sep=0pt},
  lead/.style={gray!55, line width=0.5pt},
  uni/.style={gray!45, line width=0.6pt, rounded corners=2.4pt,
              dash pattern=on 1.2mm off 0.9mm},
]

% =========================================================================
% PANEL (a) — any two frontier peers
% =========================================================================
\draw[uni] (0.05,3.85) rectangle (8.55,6.35);
\node[kick, anchor=south west] at (0.05,6.45)
  {{\upshape\bfseries (a)}~any two frontier peers};
\node[micro, anchor=north east] at (8.40,6.22) {all tasks};

\def\PA{(4.07,4.85)} \def\PB{(4.53,4.85)} \def\PR{(1.90 and 0.86)}

\fill[cBase!36] \PA ellipse \PR;
\fill[cBase!36] \PB ellipse \PR;
\begin{scope}
  \clip \PA ellipse \PR;
  \fill[cBase!12] \PB ellipse \PR;
\end{scope}
\draw[gray!62, line width=0.7pt] \PA ellipse \PR;
\draw[gray!62, line width=0.7pt] \PB ellipse \PR;

\node[lbl] at (4.30,4.78) {solved by both};

% one shared callout for the two thin crescents
\node[micro, text width=2.6cm] at (4.30,6.00)
  {solved by only one:\\rare};
\draw[lead] (3.16,5.88) .. controls (2.72,5.58) and (2.50,5.34) .. (2.40,5.02);
\draw[lead] (5.44,5.88) .. controls (5.88,5.58) and (6.10,5.34) .. (6.20,5.02);

% =========================================================================
% PANEL (b) — a cheaper model vs the strongest
% =========================================================================
\draw[uni] (0.05,0.05) rectangle (8.55,3.15);
\node[kick, anchor=south west] at (0.05,3.25)
  {{\upshape\bfseries (b)}~a cheaper model vs.\ the strongest};

\def\SC{(3.45,1.40)} \def\SR{(2.40 and 1.05)}
\def\CC{(4.80,1.32)} \def\CR{(1.26 and 0.72)}

% 1. cheap set underneath — what stays visible is the sliver outside
\fill[cBase!36] \CC ellipse \CR;
% 2. strong set on top — covers everything it contains
\fill[cBase!12] \SC ellipse \SR;
% 3. the intersection, the region this paper acts on
\begin{scope}
  \clip \SC ellipse \SR;
  \fill[cSys!13] \CC ellipse \CR;
\end{scope}

\draw[cBase!75, line width=0.8pt] \SC ellipse \SR;
\draw[gray!62, line width=0.6pt]  \CC ellipse \CR;

\node[lbl, text width=2.0cm] at (2.10,1.72)
  {Strongest model's\\solve set};

\node[lbl, text width=2.2cm] at (4.85,2.80)
  {Cheaper model's\\solve set};
\draw[lead] (4.85,2.51) -- (4.82,2.09);

\node[micro, text=cSys!72!black, text width=2.1cm] at (4.78,1.28)
  {a cheaper model\\suffices here:\\\textbf{route for cost}};

\node[micro, anchor=west, text width=1.85cm] at (6.42,1.88)
  {solved only by the\\cheaper model:\\rare};
\draw[lead] (6.36,1.64) .. controls (6.18,1.56) and (6.06,1.50) .. (5.96,1.40);

\node[micro, anchor=west] at (0.24,0.32) {solved by neither};

\end{tikzpicture}}
\caption{\textbf{Solve sets are nearly nested (schematic).} (a)~Even two
frontier peers solve almost the same tasks, so accuracy routing has little
to win. (b)~A cheaper model's set sits mostly inside the strongest
model's; that shared region is where routing for cost pays.}
\label{fig:solve-sets}
\end{figure}
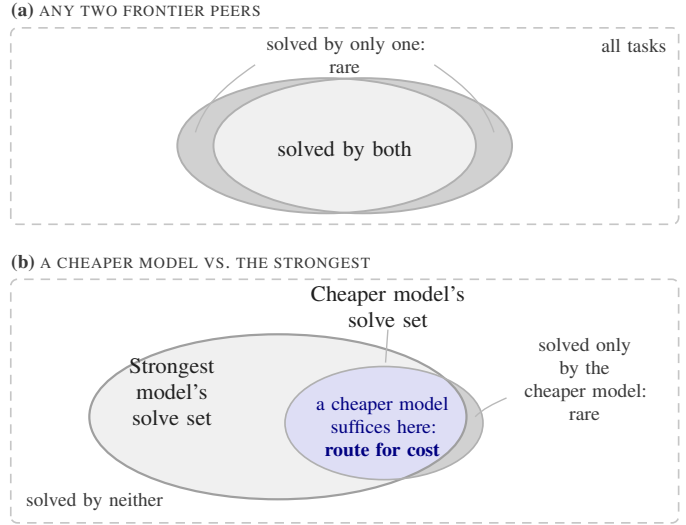

First, solve sets are strongly nested: the tasks a weaker model solves
are largely a subset of those the strongest model solves.  Measured as set
containment, the overlap is 0.941, 0.912, and 0.773 across the three
benchmarks.  Nor is this an artifact of comparing cheap models with
frontier ones: replaying pools built from frontier models alone, one per
lab, leaves containment at 0.90 to 0.93.  These models are generally
strong rather than complementary specialists, so an accuracy-oriented
router has almost no room to combine their strengths.
Figure~\ref{fig:solve-sets} sketches the consequence: between any two
models the accuracy prize is only a thin sliver, while the large shared
region is where a cheaper model would have sufficed all along.

Second, no learned router we tested exceeded the solve rate of
always picking the strongest model; every observed gap fell within noise.
Taken together, these results suggest a plain conclusion: routing for
\emph{accuracy} has little headroom on current issue-resolution
benchmarks.  Routing for \emph{cost}, by contrast, does not require
complementary skills at all.  It requires only the ability to predict when
a cheaper model will suffice.  Full audit tables appear in Appendix~\ref{apx:audit}.

\subsection{Goal}
\label{sec:problem:goal}

Our objective inverts the premise of prior routing work.  Rather than
seeking accuracy gains by combining models, we aim to match the strongest
available model's solve rate at a materially lower total cost per solve.

The thesis is that routing should happen \emph{after} engaging with the
problem, not from the task text alone.  In \sysname a small searcher
model first explores the repository; both its explicit work product (a
structured handoff) and its internal representation of the task (hidden
states) then inform the routing decision.
The evaluation partly supports this thesis: the engagement's payoff
arrives through the handoff more than through the routing decision
(\S\ref{sec:results}).
\S\ref{sec:system} details the architecture.

% Section 4, System.

% ---- full-width pipeline figure (the paper's one two-column float) ---------
\begin{figure*}[!t]
\centering
\input{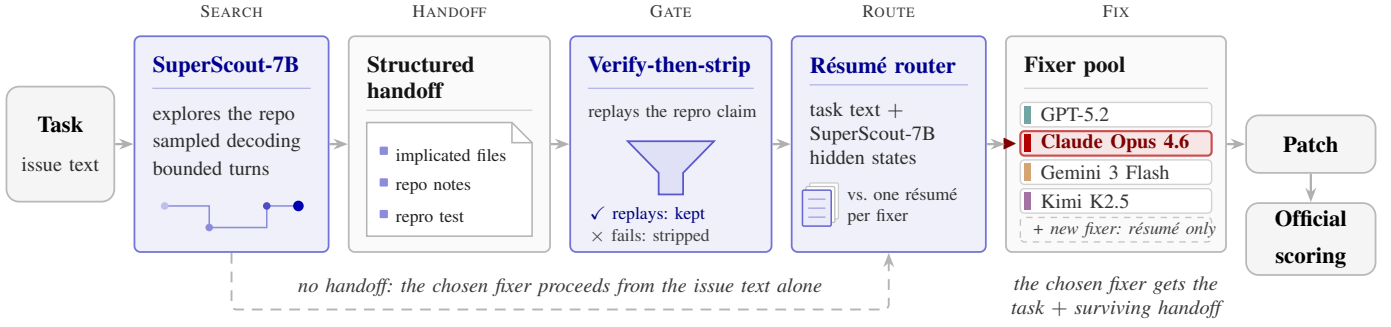}
\caption{\textbf{The \sysname pipeline.}
\modelname explores the repository and emits a structured handoff, which a
sandbox gate verifies before a r\'esum\'e router selects one of four frontier
fixers.  The dashed path is the fallback: when no handoff is produced, the
chosen fixer proceeds from the issue text alone.  Adding a new fixer requires
only a r\'esum\'e, not retraining.}
\label{fig:pipeline}
\end{figure*}

\section{The \sysname System}
\label{sec:system}

Figure~\ref{fig:pipeline} traces a single task through the pipeline.  The task
enters a search phase, where \modelname explores the repository and writes a
structured handoff.  A sandbox gate then verifies the handoff's reproduction
claims before a r\'esum\'e router selects one fixer from a pool of frontier
models.  That fixer receives the task and the surviving handoff, produces a
patch, and submits it to official scoring.  The entire trajectory is a single
pass: no parallel sampling, no cross-model escalation mid-task.  When the
searcher produces no handoff at all, the chosen fixer proceeds from the issue
text alone; this fallback defines the floor as fixer-solo performance.

\subsection{The Searcher and Its Handoff}
\label{sec:searcher}

\modelname is a 7B model built on Qwen2.5-Coder~\citep{qwen25coder}, trained
exclusively for the search phase (\S\ref{sec:training}).  Given a task, it
explores the repository, localizes the implicated files, attempts to write a
failing reproduction test, and then produces a handoff document before stopping.
The handoff is a structured artifact kept short by design, typically
about a page of text (4\,KB).  It contains implicated
files with line regions, ranked by confidence; a reproduction attempt specifying
a file, command, and observed output; dead ends the searcher already tried; and
free-form repository notes.  A verified example appears in
Appendix~\ref{apx:handoff-example}.

When the search exhausts its turn budget without committing to a handoff, a
single extra generation step demands one.  Such handoffs are tagged
\emph{forced} and are never pooled with spontaneous ones anywhere in this paper.

\subsection{Verify-then-Strip}
\label{sec:verify-strip}

The searcher's reproduction claims are not trusted.  Before any fixer sees a
handoff, a sandbox replays the claimed reproduction command against the
unpatched repository.  A claim that does not genuinely fail is stripped: both
the test file and the claim itself are removed from the handoff.  A verified
claim, by contrast, is materialized so the fixer can use it directly.
Calibration revealed that most reproduction claims are in fact false
(\S\ref{sec:calibration}), and \S\ref{sec:results} quantifies the guard's
effect at benchmark scale.  Injection is blanket: every routed fixer receives
the surviving handoff identically.

\subsection{The R\'esum\'e Router}
\label{sec:router}

\begin{table}[!tbp]
\centering
\caption{\textbf{The r\'esum\'e router, end to end.}  Each fixer's
r\'esum\'e stores two embedding centroids and a base rate, computed from
public per-task outcomes.  A logistic head per feature space scores
$P(\text{solve})$; the router walks the pool in cheap-first order,
stopping at the first fixer clearing~$\theta$.  Adding a new fixer
requires only 25--50 public outcomes and no retraining.}
\label{tab:router-recipe}
% ===========================================================================
% tab-router-recipe.tex — the deployed resume router, end to end (§4).
% BODY ONLY (tabular). Wrapper/caption/label live in the section file.
% Provenance: code/router/router_spec.json (embedder pin, resume definition,
% feature recipe, hidden-state key, head, route rule and cheap order).
%
% Notes for editing:
%   - The embedder is frozen infrastructure: one exact model and revision
%     computes every embedding in the system and is never fine-tuned.
%   - Features are listed uncentered and are computed per feature space.
%   - Wording rule: phrase onboarding as "adding a new FIXER requires no
%     retraining", never a bare "without retraining".
% ===========================================================================
\begin{tabular}{@{}ll@{}}
\toprule
Component & Specification \\
\midrule
Embedder      & Qwen3-Embedding-0.6B \\
              & frozen \\
              & 4096-token cap, $L_2$-normalized \\
\addlinespace
R\'esum\'e     & per fixer: mean embedding of \\
              & solved, mean of failed, base rate \\
              & from 25--50 public outcomes \\
\addlinespace
Features      & $\cos(x,s)$, $\cos(x,f)$, \\
              & $\cos(x,s){-}\cos(x,f)$, $p$, \\
              & $\cos(x,s){\cdot}p$, \\
              & $(\cos(x,s){-}\cos(x,f)){\cdot}p$ \\
              & uncentered, per feature space \\
\addlinespace
Hidden state  & \modelname\ layer $-4$, pre-decode \\
              & final-position state (3584-d) \\
\addlinespace
Scorer        & logistic regression per space, \\
              & uniform probability average, \\
              & $\theta{=}0.30$, cheapest-adequate \\
\addlinespace
New fixer     & 25--50 public outcomes; \\
              & no retraining of \modelname, \\
              & the embedder, or the scorer \\
\bottomrule
\end{tabular}

\end{table}

Each fixer in the pool is summarized by a r\'esum\'e built from 25--50 public
per-task outcomes.  A r\'esum\'e stores three quantities: the mean embedding of
tasks the fixer solved, the mean embedding of tasks it failed, and a base solve
rate.  Table~\ref{tab:router-recipe} lists the full specification.

At routing time the task is embedded in two complementary feature spaces.  Its
text is encoded by a frozen off-the-shelf embedder~\citep{qwen3embedding}.  Its
content as \modelname experienced it is captured through the searcher's
3{,}584-dimensional pre-decode hidden state, drawn from the fourth-from-last
layer at the final token position.  In each feature space the router computes
uncentered cosine similarities between the task and every r\'esum\'e's solved
and failed centroids, their difference, and the base rate; a single logistic
regression per feature space, shared across fixers, then scores
$P(\text{solve})$ for each fixer from its r\'esum\'e-relative features.  The
routing head blends the
task-text and hidden-state feature sets by uniformly averaging their predicted
probabilities.

Routing walks the pool in cheap-first order and assigns the task to the first
fixer whose predicted probability clears a caution threshold~$\theta$; if none
clears it, the task falls to a designated anchor model (\S\ref{sec:setup}).  Logistic regression was chosen
over a multi-layer perceptron during calibration, where the simpler scorer won
repeatedly (\S\ref{sec:calibration}).

\FloatBarrier

\subsection{Extensibility}
\label{sec:extensibility}

Adding a new fixer requires no retraining.  A new model's r\'esum\'e consists
of two averaged embeddings and a base rate, computed from its public outcomes.
The embedder is frozen and \modelname is untouched; \S\ref{sec:results}
presents a measured discussion.

% Section 5, Training the Searcher.
\section{Training the Searcher}
\label{sec:training}

\subsection{Data and Supervised Training}
\label{sec:training-data}

\modelname is trained on search-phase demonstrations sliced from openly licensed
agent trajectories produced by other systems. Each demonstration captures one
complete search episode: the agent explores a repository, localizes the fault,
and writes a failing reproduction, with a synthesized handoff-emission turn
appended as the final supervised target. Within that synthesized turn, the
file list is extracted deterministically from the trace and the reproduction
record is copied verbatim; only the free-form notes are written by a model,
the open-weights gpt-oss-120b~\citep{gptoss}. Traces are drawn from three public
sources (\dataset{Open-SWE-Traces}~\citep{openswetraces},
\dataset{SWE-rebench-openhands}~\citep{nebiusrebench,swerebench}, and
\dataset{SWE-Hero}~\citep{swehero}; the latter two are trajectories of
the OpenHands scaffold~\citep{openhands}) and then success-filtered
against the gold patch, keeping only trajectories that actually found the right files.
Deduplication retains the two highest quality-ranked traces per issue, so the
corpus counts distinct bugs rather than retellings of the same fix.

The resulting dataset contains 19{,}911 examples. Six rows carrying malformed
tool-call wrappers are dropped at load time, giving the 19{,}905 examples
actually trained on. The realized language mix is Python 37.3\%, Go 36.7\%,
TypeScript 19.7\%, and JavaScript 6.3\%.
Table~\ref{tab:training-data} summarizes the corpus composition.

\begin{table}[!tbp]
\centering
\caption{\textbf{Composition of \modelname's supervised search corpus.}
Each example is a full search trajectory with a synthesized
handoff-emission turn as the final supervised step; loss is on assistant
turns only.  Traces are success-filtered and deduplicated to two per
issue.  The 23 evaluation repositories and 450 held-out vault issues are
excluded; both exclusions were verified by a programmatic gate.}
\label{tab:training-data}
\begin{tabular}{@{}lr@{}}
\toprule
Component & Value \\
\midrule
\multicolumn{2}{@{}l}{\emph{Corpus}} \\
Examples trained on          & 19,905 \\
Built / frozen set           & 19,911 \\
Held-back shelf (untrained)  & 9,478 \\
Tokens per epoch             & 382.4M \\
Packed 32k blocks (fill)     & 12,723 (91.7\%) \\
\midrule
\multicolumn{2}{@{}l}{\emph{Language mix}} \\
Go                           & 7,304 (36.7\%) \\
Python                       & 7,417 (37.3\%) \\
TypeScript                   & 3,932 (19.7\%) \\
JavaScript                   & 1,258 \phantom{0}(6.3\%) \\
\midrule
\multicolumn{2}{@{}l}{\emph{Trace sources}} \\
Open-SWE-Traces              & Go/TS/JS + Py \\
SWE-rebench-openhands        & Python \\
SWE-Hero-openhands           & Python \\
\midrule
\multicolumn{2}{@{}l}{\emph{Filters}} \\
Traces kept per issue        & top 2 \\
Verified reproduction        & 97.7\% \\
Excluded repositories        & 23 \\
Excluded held-out issues     & 450 \\
\bottomrule
\end{tabular}

\end{table}

Contamination control was decided on day one: a 23-repository blocklist
covering all 11 SWE-bench Pro repositories and 12 SWE-bench
Verified~\citep{swebenchverified} repositories is excluded from every
training and calibration set. A
programmatic gate verified zero blocklist hits in the frozen file.

We fine-tune Qwen2.5-Coder-7B~\citep{qwen25coder} with
LoRA~\citep{lora} on a single GPU. Before training, the base model's
localization rate is near zero; the search behavior is entirely learned from
the demonstrations above. Hyperparameters are reported in Appendix~\ref{apx:hyperparams}.

\subsection{Decoding Matters More Than Expected}
\label{sec:decoding}

The most consequential training-era finding was not about the training itself
but about inference. On a 450-task held-out exam, greedy decoding finds the
right files at a rate of 0.1104. A single sampled draw at temperature 0.9
reaches 0.3058, a 2.65$\times$ gain after matching for infrastructure timeouts.

\begin{figure}[!tbp]
% ===========================================================================
% fig-decoding.tex — the sampled-decoding discovery, two panels.
% BODY ONLY (tikzpicture). Wrapper/caption/label live in the section file.
% Requires \input{figures/fig-style} in the preamble.
%
% PROVENANCE — every number.
%   PANEL 1 (450-task vault). Source: internal records, header table
%     "450-VAULT MILESTONE — merged-e2.0 —
%     run A (temp 0.9) vs run B (greedy)", row `find_rate`:
%       RUN B (greedy)   0.1104
%       RUN A (temp 0.9) 0.3058
%     n = 450 per arm (row `n`).
%     Multiplier: the same source's timeout-sensitivity section —
%       RAW      A=0.3058 B=0.1104  ratio 2.77x
%       MATCHED  A=0.3091 B=0.1166  ratio 2.65x  (n=426)
%     The LICENSED multiplier is the timeout-matched 2.65x, which is what is
%     annotated on the panel.
%     *** Never quote a 5.7x multiplier anywhere: that figure's denominator
%     was a prefix-cache-contaminated run. ***
%
%   PANEL 2 (100-task dial subset — a DIFFERENT task set from panel 1).
%     Source: internal records, the decomposition of where the gain comes
%     from:
%       "commit rate 3.03x  x  per-handoff quality 0.96x  =  find-rate 2.91x"
%     Underlying per-handoff recall: greedy 0.526 vs temp 0.9 0.505 (same
%     section); commit/handoff coverage 0.2367 -> 0.7167 (CHECK 2 table).
%     Task set = dial_subset_100 (md5 cb363beb...), 300 episodes per arm
%     (3 replicates x 100), per CHECK 1 / CHECK 2 tables.
% ===========================================================================
\begin{tikzpicture}
\begin{groupplot}[
  paperaxis,
  group style={
    group size=2 by 1,
    horizontal sep=13mm,
  },
  width=0.44\linewidth,
  height=5.0cm,
  xticklabel style={font=\footnotesize},
  ymajorgrids=true,
  enlarge x limits=0.55,
  nodes near coords,
  nodes near coords style={font=\scriptsize, color=black},
]

% ---- PANEL 1: find rate, vault-450 -------------------------------------
\nextgroupplot[
  ybar, bar width=11pt,
  title={Find rate (vault-450)},
  ylabel={Mean gold-file recall},
  ymin=0, ymax=0.40,
  ytick={0,0.1,0.2,0.3,0.4},
  symbolic x coords={greedy,{temp 0.9}},
  xtick={greedy,{temp 0.9}},
  nodes near coords={\pgfmathprintnumber[fixed, fixed zerofill, precision=3]{\pgfplotspointmeta}},
]
\addplot[fill=cBase!45, draw=cBase!80!black, bar shift=0pt]
  coordinates {(greedy,0.1104)};
\addplot[fill=cSys!35, draw=cSys, bar shift=0pt,
         postaction={pattern=north east lines, pattern color=cSys!45}]
  coordinates {({temp 0.9},0.3058)};
\node[font=\scriptsize, anchor=south, align=center, text=cSys!80!black]
  at (axis cs:greedy,0.20) {$2.65\times$\\timeout-\\matched};

% ---- PANEL 2: decomposition, 100-task dial subset ----------------------
\nextgroupplot[
  ybar, bar width=11pt,
  title={Decomposition (dial-100)},
  ylabel={Ratio, temp 0.9 $/$ greedy},
  ymin=0, ymax=3.9,
  ytick={0,1,2,3},
  symbolic x coords={{commit rate},{per-handoff}},
  xtick={{commit rate},{per-handoff}},
  xticklabels={commit\\rate, handoff\\quality},
  xticklabel style={font=\scriptsize, align=center},
  nodes near coords={$\pgfmathprintnumber[fixed, precision=2]{\pgfplotspointmeta}\times$},
]
\addplot[fill=cSys!35, draw=cSys, bar shift=0pt]
  coordinates {({commit rate},3.03)};
\addplot[fill=cBase!45, draw=cBase!80!black, bar shift=0pt]
  coordinates {({per-handoff},0.96)};
% parity line at ratio 1.0; ymin=0, ymax=3.9 -> axis fraction 1/3.9.
\draw[dashed, gray!70, line width=0.6pt]
  (rel axis cs:0,0.25641) -- (rel axis cs:1,0.25641);
\node[font=\scriptsize, anchor=south west, text=gray!60!black]
  at (rel axis cs:0.03,0.29) {parity};

\end{groupplot}
\end{tikzpicture}
\caption{\textbf{Sampled decoding is what makes the searcher useful.}
\textbf{Left:} switching from greedy to temperature $0.9$ raises the
find rate from $0.110$ to $0.306$ ($2.65\times$, $n{=}426$ of $450$).
\textbf{Right:} on a separate 100-task dial set, commitment rises
$3.03\times$ while per-handoff quality holds at $0.96\times$; on the
exam itself emission rises $3.4\times$ against an 18\% per-handoff
recall cost (Appendix~\ref{apx:vault}).  The two panels use different
task sets; the right panel is a mechanism check, not a replication.}
\label{fig:decoding}
\end{figure}
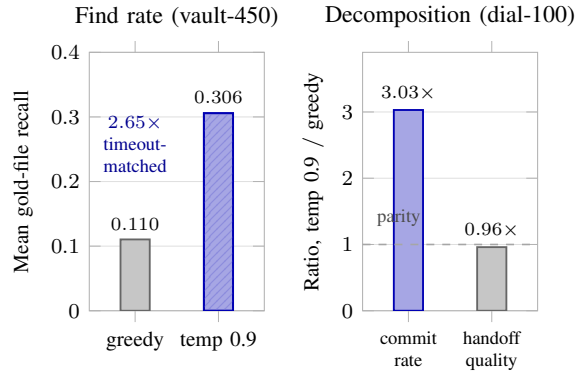

The exam's own decomposition shows what sampling buys and what it
spends: the searcher's emission rate rises from 0.213 to 0.718
(3.4$\times$), while recall per emitted handoff falls from 0.517 to
0.426, an 18\% quality cost (Appendix~\ref{apx:vault}). Sampling trades
a slice of per-handoff quality for a much larger gain in willingness to
commit, a strongly net-positive exchange. A separate 100-task dial set
shows the same mechanism with commitment up 3.03$\times$ and per-handoff
quality flat at 0.96$\times$ (Figure~\ref{fig:decoding}); the quality
cost visible on the exam does not appear there, so the dial set is a
mechanism check, not a replication. Greedy decoding, it turns out,
makes the searcher reluctant to declare a result; sampling at moderate
temperature recovers ability the model already has. The shipped
configuration is therefore a single sampled draw at temperature 0.9
(\S\ref{sec:setup}).

\subsection{Language Transfer}
\label{sec:language-transfer}

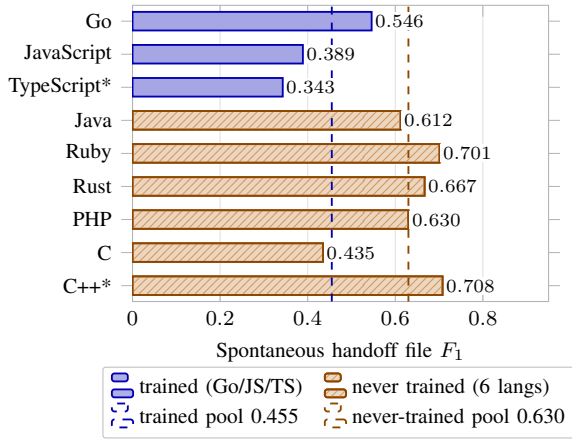
\begin{figure}[!tbp]
\begin{tikzpicture}
\begin{axis}[
  paperaxis,
  width=0.80\linewidth,
  height=5.5cm,
  xbar,
  bar width=7pt,
  xmin=0, xmax=0.95,
  xtick={0,0.2,0.4,0.6,0.8},
  xlabel={Spontaneous handoff file $F_1$},
  symbolic y coords={{Go},{JavaScript},{TypeScript*},
                     {Java},{Ruby},{Rust},{PHP},{C},{C++*}},
  ytick={{Go},{JavaScript},{TypeScript*},
         {Java},{Ruby},{Rust},{PHP},{C},{C++*}},
  y dir=reverse,
  enlarge y limits=0.06,
  xmajorgrids=true,
  nodes near coords,
  % white fill so the two dashed pool lines never run through a value label
  nodes near coords style={font=\scriptsize, fill=white, inner sep=1pt,
                           /pgf/number format/fixed,
                           /pgf/number format/fixed zerofill,
                           /pgf/number format/precision=3},
  legend style={at={(0.5,-0.22)}, anchor=north, legend columns=2,
                /tikz/every even column/.append style={column sep=6pt}},
  legend cell align=left,
]

% ---- trained pool: Go / JavaScript / TypeScript (solid) ----------------
\addplot[fill=cSys!35, draw=cSys, bar shift=0pt]
  coordinates {(0.546,{Go}) (0.389,{JavaScript}) (0.343,{TypeScript*})};
\addlegendentry{trained (Go/JS/TS)}

% ---- never-trained pool (hatched, so the split survives grayscale) -----
\addplot[fill=cFlash!25, draw=cFlash!80!black, bar shift=0pt,
         postaction={pattern=north east lines, pattern color=cFlash!60}]
  coordinates {(0.612,{Java}) (0.701,{Ruby}) (0.667,{Rust})
               (0.630,{PHP}) (0.435,{C}) (0.708,{C++*})};
\addlegendentry{never trained (6 langs)}

% ---- pooled aggregates: 0.455 (trained) and 0.630 (never-trained) ------
% x positions as axis fractions: 0.455/0.95 and 0.630/0.95.
\draw[dashed, cSys, line width=0.7pt]
  (rel axis cs:0.47895,0) -- (rel axis cs:0.47895,1);
\draw[dashed, cFlash!80!black, line width=0.7pt]
  (rel axis cs:0.66316,0) -- (rel axis cs:0.66316,1);
% the two dashed lines are identified in the legend rather than in-plot,
% where they would collide with the bar value labels.
\addlegendimage{dashed, cSys, line width=0.7pt}
\addlegendentry{trained pool 0.455}
\addlegendimage{dashed, cFlash!80!black, line width=0.7pt}
\addlegendentry{never-trained pool 0.630}

\end{axis}
\end{tikzpicture}
\caption{\textbf{Localization transfers to languages \modelname was never
trained on.}  Per-language file-level $F_1$ of spontaneous handoffs
(searcher alone, single draw, SWE-bench Multilingual).  Solid bars:
three trained languages; hatched: six never-trained.  Dashed lines mark
pooled aggregates ($0.455$ trained, $0.630$ never-trained).
The TypeScript and C++ cells (12 assigned tasks each; 7 and 8
spontaneous handoffs analyzed) are indicative only.}
\label{fig:language-transfer}
\end{figure}

\modelname is trained on four languages (Go, Python, TypeScript, JavaScript),
yet \sysname is deployed against repositories in languages the searcher has
never seen. On the nine-language SWE-bench Multilingual
evaluation~\citep{swebenchmultilingual}, which contains no Python and so
covers three of the four training languages
(Figure~\ref{fig:language-transfer}), spontaneous-handoff localization quality
is higher on the six never-trained languages (file-level $F_1 = 0.630$) than on
the three trained ones ($F_1 = 0.455$). The inversion survives a
difficulty-matched control restricted to single-gold-file tasks, where recall
on never-trained languages reaches 0.791 versus 0.524 for the trained pool.
JavaScript and TypeScript account for much of the trained pool's lower average;
the TypeScript and C++ cells each rest on 12 assigned tasks, of which 7 and 8
respectively produced the spontaneous handoffs analyzed, and are indicative
only (full per-language table in Appendix~\ref{apx:languages}).

The contrast carries a confound: JavaScript and TypeScript are at once the
trained pool's weakest cells and its smallest training slices, and both may be
structurally harder to localize in (dynamic imports, build artifacts)
independently of training exposure, so the trained-versus-never-trained
comparison should not be read as causal.
Our hypothesis for the pattern, and it is a hypothesis, is that \modelname
learned a search method, not a language-specific vocabulary. The exploration
strategies it relies on (reading directory trees, tracing imports, scanning
test files) are structural operations that generalize across languages; the
specific tokens involved matter less than the strategy of following them.

\medskip
\noindent\textbf{Reinforcement learning.}\quad
We built and validated a GRPO~\citep{grpo} training rig and ran 50 clean
steps. Learning was flat. A trace-level autopsy showed that the reward signal
was real but aimed at a behavior already near its ceiling after supervised
training: the model could localize when it chose to commit, and sampling had
already recovered that commitment. We shelved RL in favor of the decoding fix
described above; rig details are in Appendix~\ref{apx:hyperparams}.

\FloatBarrier

% Section 6, Experimental Setup.
\section{Experimental Setup}
\label{sec:setup}

\subsection{Benchmark and Protocol}
\label{sec:setup-bench}

We evaluate on SWE-bench Pro~\citep{swebenchpro}, using the full Python
slice: all 266 tasks, a census rather than a sample.  These span three
repositories: 96 from ansible, 91 from openlibrary, and 79 from
qutebrowser, out of 731 tasks in the public set.  We chose Pro because
its tasks are harder and less saturated than earlier SWE-bench variants,
and because all 11 of its repositories were excluded from \modelname's
training data from day one via the contamination blocklist of
\S\ref{sec:training}.

Our protocol matches the benchmark's official capped budget tier
exactly: a 50-LLM-call turn cap and a \$2.00 per-attempt cost limit.
The cap is silent; the agent is never told a budget exists.
Source-code inspection of the official scaffold found no budget
sentence in any prompt; when the cap is exhausted, the current working
diff is auto-submitted.  Task input follows the scaffold's exact
three-field concatenation of problem statement, requirements, and
interface.  For 105/266 tasks a byte-lossless JSON decode of
double-serialized fields was applied; this deviation is logged.  Fixer
prompts are byte-identical to the SWE-bench Multilingual leaderboard
templates~\citep{swebenchmultilingual}, and a single task-text
definition is used throughout, for the searcher, every fixer, and the
router embedding alike.

\subsection{Frozen System Configuration}
\label{sec:setup-frozen}

The searcher, gate, r\'esum\'es, embedder, and router weights were
frozen before first contact with the benchmark.  \modelname serves at
temperature~0.9, single draw, with pinned sampling parameters,
per-episode cache isolation, a serving-health canary gate, and a 40-turn
search budget.  The handoff policy is blanket injection: every routed
fixer receives the surviving handoff identically.  Verify-then-strip
runs first on every reproduction claim (mechanism in
\S\ref{sec:system}).  The router's logistic head was fit on all
calibration data (recipe in Table~\ref{tab:router-recipe}) and operates
at caution threshold $\theta{=}0.30$.  This threshold was initially
calibrated against solo-fixer outcomes; because the handoff
systematically lifts inexpensive fixers, it was recalibrated on the
99-task lab set using with-handoff outcomes via the pre-existing
matched-point rule.  The router's benchmark result is evaluated by
routing each task and scoring it against that fixer's measured episode
under the official capped protocol; every one of the 266 routed picks
has a real measured episode behind it.

\subsection{Arms, Baselines, and Cost Accounting}
\label{sec:setup-arms}

Four measured arms run fully paired on the same 266 tasks: three solo
fixers (\fxopus~\citep{opus46}, \fxgpt~\citep{gpt52}, and
\fxkimi~\citep{kimik25}) and the \sysname fixer runs with injected
handoffs.  Both frontier models were run in full as solo baselines
because our calibration study and a small paid probe disagreed about
which was stronger; the resolution is itself a result
(\S\ref{sec:results}).  At the calibrated operating point
($\theta{=}0.30$) the router's fallback never fires, so anchor choice is
moot.

Two baselines anchor interpretation.  The strongest-solo baseline is the
single fixer with the highest solve rate.  The blind-mixing line traces
the accuracy-versus-cost segment produced by randomly assigning each
task to \fxkimi or \fxopus in varying proportions; a system point above
this line does something smarter than chance mixing.

Cost accounting is deliberately conservative against \sysname.  System
costs are all-in: routed-fixer API spend plus \modelname's entire GPU
bill plus verify-then-strip sandbox infrastructure, all amortized per
task.  Solo arms count only their pure fixer API spend.  We report total
cost per solve at a matched solve rate; measured spend is disclosed in
Appendix~\ref{apx:spend}.  Scoring uses the benchmark's
official harness; a gold-patch control passes 265/266 tasks.  The single
failure stems from a dataset artifact (a truncated test-spec
identifier) that is unwinnable for every arm equally and is kept in all
denominators.

% Section 7, Results.
\section{Results}
\label{sec:results}

% ---------------------------------------------------------------------------
% 7.1  Headline
% ---------------------------------------------------------------------------
\subsection{Headline}
\label{sec:results-headline}

\begin{table}[!tbp]
\centering
\caption{\textbf{Main results on SWE-bench Pro (Python-266).}
\sysname matches the pool's best single model (\fxopus) at about a fifth
of its cost per solve; 159 versus 158 is a one-task gap at $n{=}266$, a
match, not a win.  The last row is the no-router ablation, which ties
the routed system on this benchmark (\S\ref{sec:results-headline}).}
\label{tab:main-results}
% ===========================================================================
% tab-main-results.tex — main results, SWE-bench Pro Python-266.
% BODY ONLY (tabular): the table wrapper, \caption and \label live in the
% section file that includes this.
% Provenance: data/receipts/outcomes_per_task.json.
%
% "vs. mix" = percentage points above the blind-mixing line, i.e. the linear
%   interpolation between the two solo endpoints, \fxkimi ($0.106138/task,
%   56.0150%) -> \fxopus ($0.756586/task, 59.3985%) in ($/task, solve-rate)
%   space, evaluated at the row's own $/task; row rates use the exact
%   fraction 159/266, not the rounded 59.77%.
% Bolding: best value per column; genuine ties are bolded on both rows.
%   Wording rule: "matches", never "beats" — 159 vs 158 is one task at n=266.
% Layout: wide table. House rule is a smaller font plus tighter \tabcolsep
%   kept local to this tabular, never \resizebox. The first column is a
%   fixed-width paragraph column so the two long system labels wrap instead
%   of pushing the tabular past the text column.
% ===========================================================================
\begingroup\footnotesize\setlength{\tabcolsep}{2pt}%
\begin{tabular}{@{}p{1.20in}rrrrr@{}}
\toprule
System & Solves/266 & Rate & \$/task & \$/solve & vs.\ mix \\
\midrule
\raggedright\sysname (router, $\theta{=}0.30$)
  & \textbf{159} & \textbf{59.77\%} & \$0.137 & \$0.230 & $+3.60$ \\
\midrule
\fxopus solo  & 158 & 59.40\% & \$0.757 & \$1.274 & --- \\
\fxkimi solo  & 149 & 56.02\% & \textbf{\$0.106} & \textbf{\$0.190} & --- \\
\fxgpt solo   & 139 & 52.26\% & \$0.570 & \$1.091 & --- \\
\midrule
\raggedright No-router ablation (\fxkimi $+$ handoff)
  & \textbf{159} & \textbf{59.77\%} & \$0.136 & \$0.227 & $+3.60$ \\
\bottomrule
\multicolumn{6}{@{}p{0.97\linewidth}@{}}{\footnotesize Solves are out of
$n{=}266$, top to bottom: 159, 158, 149, 139, 159. The \sysname row and the
ablation row share one all-in cost convention: fixer API spend plus the
\modelname searcher's GPU bill plus verify-then-strip sandbox infrastructure,
amortized over 266 tasks. Solo rows are fixer API spend only. ``vs.\ mix'' is
percentage points above the \fxkimi--\fxopus blind-mixing line at the row's
own \$/task.}
\end{tabular}%
\endgroup

\end{table}

Table~\ref{tab:main-results} summarizes the primary evaluation.
\sysname solves 159 of 266 tasks (59.77\%) at a total cost per solve of
\$0.230.  The pool's best single model, \fxopus run solo, solves 158 of
266 (59.40\%) at \$1.274 per solve.  \sysname matches this ceiling at
about a fifth of the cost per solve.  The one-task gap between 159 and
158 at $n{=}266$ is not a meaningful difference; it is a match.

The no-router ablation sends every task to \fxkimi with the handoff and
no routing decision at all, also solving 159 of 266 at \$0.227 per
solve.  The calibrated router concentrates 263 of 266 tasks on the
cheapest fixer and adds no solves over this ablation; its three
diversions to \fxflash~\citep{gemini3flash} cost \$0.003 per solve,
insurance rather than accuracy.  On this benchmark, then, the handoff
carries the result and routing collapses to cost allocation.  Whether
blanket assignment to one cheap fixer reaches parity is a property of
this pool and benchmark, not a general rule: in deployment that answer
is not known in advance, and the router is the component that predicts
it, from free public data, before any spend.  The ablation is ex-ante
specifiable too, but only hindsight shows it reaches parity; the router
turns that gamble into a calibrated decision.  Both
system rows sit above the blind-mixing line defined in
\S\ref{sec:setup} by $+$3.60 percentage points at their respective cost
points (visible in Figure~\ref{fig:cost-solve}).  \modelname's entire
GPU bill for all 266 search episodes was \$1.13, under half a cent per
task.

Why were both frontier models run in full as solo baselines?  Our
calibration study and a small paid probe disagreed about which was
stronger.  On the fresh calibration tasks (\S\ref{sec:calibration}),
\fxgpt was the strongest solo fixer at 60.6\% solve rate.  A small paid
probe of 15 tasks per fixer leaned the other way: \fxopus solved two
more tasks, but $p{=}0.625$, well below the pre-declared decision bar,
while inverting the calibration prior.  Both were therefore run in full.
The complete paired comparison resolved the question: \fxopus beats
\fxgpt by $+$7.14 percentage points on the same 266 tasks, with 44
tasks solved only by \fxopus versus 25 solved only by \fxgpt (exact
McNemar test~\citep{mcnemar}, $p{=}0.029$).

One further observation from the solo baselines deserves note.
\fxkimi solo resolves 149 of 266 tasks (56.02\%) at \$0.106 per task,
dominating \fxgpt solo (139/266, 52.26\%, \$0.570 per task) on both
axes, gaining $+$3.76 percentage points of accuracy at roughly one fifth
of the per-task cost.

% ---------------------------------------------------------------------------
% 7.2  Component Analysis
% ---------------------------------------------------------------------------
\subsection{Component Analysis}
\label{sec:results-components}

\paragraph{Localization quality.}

\begin{table}[!tbp]
\centering
\caption{\textbf{\modelname's file localization on SWE-bench Pro
(Python-266).}  Per-task mean recall and precision against the gold
patch's file set.  Spontaneous and forced handoffs are reported
separately.  All numbers are from a single sampled draw at
temperature~$0.9$ (no majority vote).}
\label{tab:localization}
% ===========================================================================
% tab-localization.tex — \modelname file localization quality on Pro-266.
% BODY ONLY (tabular). Wrapper/caption/label live in the section file.
%
% PROVENANCE — every cell.
%   Source: internal records.
%     overall row (n=266)      = overall: mean recall 0.566,
%                                mean precision 0.821, all-gold-hit 24.8%
%     spontaneous row (n=206)  = spontaneous vs forced,
%                                spontaneous column: 0.586 / 0.833 / 28.2%
%     forced row (n=60)        = forced column: 0.499 / 0.780 / 13.3%
%   Method (single sampled draw, temp 0.9, no majority vote; gold = b-side
%   `diff --git` paths of `gold_patch`) = the source's method and caveats.
%
%   RULE APPLIED: the per-repo breakdown (ansible / openlibrary /
%   qutebrowser) is APPENDIX-ONLY and is
%   deliberately omitted here.
%   Spontaneous and forced are NEVER pooled in the row-level breakdown; the
%   overall row pools them only for the single top-line number, per the
%   source's own rule.
% ===========================================================================
\begin{tabular}{@{}lrrr@{}}
\toprule
Handoff type & Recall & Precision & All-gold \\
\midrule
Overall ($n{=}266$)      & 0.566 & 0.821 & 24.8\% \\
\midrule
Spontaneous ($n{=}206$)  & \textbf{0.586} & \textbf{0.833} & \textbf{28.2\%} \\
Forced ($n{=}60$)        & 0.499 & 0.780 & 13.3\% \\
\bottomrule
\end{tabular}

\end{table}

Table~\ref{tab:localization} reports the quality of \modelname's file
localization.
Against a mean of 3.44 gold files per task, the searcher names a median
of 2.0 files, achieving per-task mean recall of 0.566 and precision of
0.821; in 24.8\% of tasks every gold file appears in the handoff.
Spontaneous handoffs ($n{=}206$) localize substantially better than
forced ones ($n{=}60$) across the board: recall 0.586 versus 0.499,
precision 0.833 versus 0.780, and all-gold rate 28.2\% versus 13.3\%.
This gap is unsurprising: tasks that exhaust the turn budget are
typically harder.  It underscores why the two handoff types are
never pooled when reporting component behavior (per-repository detail in Appendix~\ref{apx:localization}).

\paragraph{Verify-then-strip guard.}
Of 266 handoffs, 249 included a claimed verified reproduction of the
bug.  Replaying every claim inside the task sandbox revealed that only
50 of them (20\%) were genuine, while 174 (70\%) were demonstrably
false.  The guard stripped every false claim before any fixer saw it.
Forced handoffs overclaimed more aggressively: just 9\% of their
reproduction claims were genuine, compared with 22\% for spontaneous
handoffs.
These benchmark rates are worse than those observed during calibration
(32\% genuine, 56\% false; \S\ref{sec:calibration}), consistent with
harder tasks producing more overclaiming.
The guard neutralizes this overclaiming before any fixer sees it: 174
fixer prompts were stripped of confident misinformation that would
otherwise have been treated as ground truth (full outcome classes in
Appendix~\ref{apx:verify-strip}); we did not run a pass-through arm, so
the guard's contribution to the headline is an inference from this
claim census rather than a measurement (\S\ref{sec:limitations}).

\paragraph{Protocol findings.}
The auto-submit-at-cap mechanism rescued 24 of \fxopus's 158 solves and
33 of \fxgpt's 139.  Without it, both frontier models' solve rates would read
roughly 9--13 percentage points lower.
The 50-call cap binds often: it was reached in 36.5\% of \fxgpt
attempts, 17.3\% of \fxopus attempts, and 48\% of \fxkimi attempts.
The \$2 cost cap bound only 5 times (all \fxopus, maximum \$2.09).
The gold-control census confirmed 265 of 266 tasks
(\S\ref{sec:setup}) (full cap statistics in Appendix~\ref{apx:cap-stats}).

\paragraph{Router behaviour.}

\begin{figure}[!tbp]
\centering
% ===========================================================================
% fig-gate-probability.tex — distribution of the router's gate probability for
% the cheapest fixer, under three feature spaces.
% BODY ONLY (tikzpicture). Wrapper/caption/label live in the section file.
% Requires \input{figures/fig-style} in the preamble.
% Provenance: data/receipts/routing_scenarios.json — per-task probabilities
% read from the probability_vectors map for the cheapest fixer (first entry of
% cheap_order), three fields per task: p_state (hidden-state head, cSys solid),
% p_text (task-text head, cBase dashed) and p_blend (their uniform average, the
% routing head the gate thresholds, cOpus dotted). n = 266.
%
% Binning: 19 equal bins of width 0.05, left edges 0.05 ... 0.95, half-open
% [lo, hi) with the last bin closed; each series sums to 266. Plotted as
% const-plot outlines, coordinates (left edge, count), opened with (0.05,0)
% and closed with (1.00,0). theta = 0.30 is the only threshold drawn.
% ===========================================================================
\begin{tikzpicture}
\begin{axis}[
  paperaxis,
  width=0.80\linewidth,
  height=5.2cm,
  xmin=0.05, xmax=1.00,
  ymin=0, ymax=110,
  xtick={0.1,0.2,0.3,0.4,0.5,0.6,0.7,0.8,0.9,1.0},
  xticklabel style={/pgf/number format/fixed,
                    /pgf/number format/fixed zerofill,
                    /pgf/number format/precision=1},
  ytick={0,25,50,75,100},
  xlabel={$P(\text{solve})$ for the cheapest fixer (\fxkimi)},
  ylabel={tasks},
  ymajorgrids=true,
  clip=false,   % lets the \theta tag sit just above the top frame
  legend style={at={(0.5,-0.30)}, anchor=north, legend columns=2,
                column sep=8pt},
  legend cell align=left,
]

% ---- deployed threshold, the ONLY threshold drawn ----------------------
\addplot[gray!55!black, line width=0.7pt, dashed, forget plot]
  coordinates {(0.30,0) (0.30,110)};
\node[font=\footnotesize, text=gray!55!black, anchor=south, inner sep=1.5pt]
  at (axis cs:0.30,110) {$\theta{=}0.30$};

% ---- task-text features (widest, drawn first so it sits behind) --------
\addplot[const plot, cBase!85!black, line width=0.8pt, dashed]
  coordinates {(0.05,0)
    (0.05,2) (0.10,6) (0.15,16) (0.20,20) (0.25,22) (0.30,14)
    (0.35,18) (0.40,25) (0.45,20) (0.50,22) (0.55,17) (0.60,27)
    (0.65,19) (0.70,18) (0.75,10) (0.80,6) (0.85,3) (0.90,0)
    (0.95,1) (1.00,0)};
\addlegendentry{task-text features}

% ---- routing head (uniform average of the two feature heads) -----------
\addplot[const plot, cOpus, line width=0.9pt, densely dotted]
  coordinates {(0.05,0)
    (0.05,0) (0.10,0) (0.15,0) (0.20,0) (0.25,3) (0.30,27)
    (0.35,33) (0.40,35) (0.45,44) (0.50,35) (0.55,52) (0.60,28)
    (0.65,5) (0.70,3) (0.75,1) (0.80,0) (0.85,0) (0.90,0)
    (0.95,0) (1.00,0)};
\addlegendentry{routing head (blend)}

% ---- hidden-state features (the concentrated one, drawn last, on top) --
\addplot[const plot, cSys, line width=1.0pt]
  coordinates {(0.05,0)
    (0.05,0) (0.10,0) (0.15,0) (0.20,0) (0.25,0) (0.30,0)
    (0.35,0) (0.40,58) (0.45,96) (0.50,69) (0.55,21) (0.60,4)
    (0.65,13) (0.70,5) (0.75,0) (0.80,0) (0.85,0) (0.90,0)
    (0.95,0) (1.00,0)};
\addlegendentry{hidden-state features}

\end{axis}
\end{tikzpicture}
\caption{\textbf{Two feature spaces, two very different confidence
distributions at the routing gate ($\theta{=}0.30$).}  Histograms of
$P(\text{solve})$ for \fxkimi (cheapest fixer) across all $266$ tasks.
The task-text head spans nearly the full unit interval (sd~$0.20$),
while the hidden-state head concentrates tightly (sd~$0.07$, support
entirely above~$\theta$).  Their blend, the routing head, gates $263$ of
$266$ tasks to the cheapest fixer.}
\label{fig:gate-probability}
\end{figure}
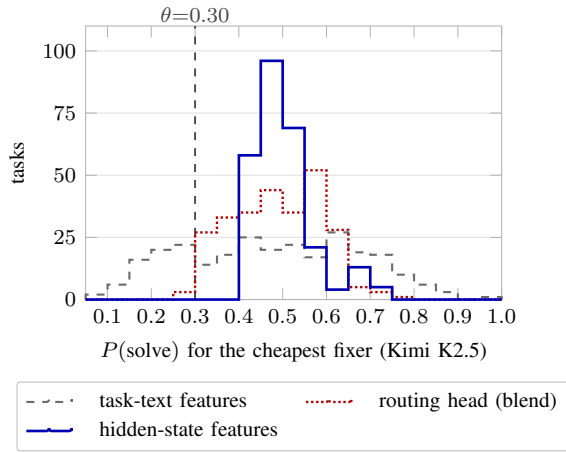

The router sends 263 of 266 tasks to \fxkimi: the first fixer in
cheap-to-expensive order whose blended solve probability clears
$\theta{=}0.30$ receives the task.
Figure~\ref{fig:gate-probability} reveals how the two underlying feature
spaces distribute that probability.  The task-text head spreads wide
(sd~$0.20$, support $0.08$--$0.97$), while the hidden-state head
concentrates tightly (sd~$0.07$, support $0.40$--$0.74$) and sits
entirely above the threshold.
Because that support lies entirely above $\theta$, part of the
hidden-state head's effect at the gate is mechanical: blending shrinks
the text head's predictions toward the state head's near-constant mean,
which by itself admits more tasks to the cheapest fixer.  We did not
run a matched shrinkage control, so the feature's informational
contribution cannot be fully separated from this threshold-shifting
effect; the held-out separation gap (AUC $0.600$ versus $0.561$ for
task text at $N{=}99$, Appendix~\ref{apx:design-space}) is suggestive
rather than decisive.
The hidden state's contribution is not boosting accuracy per se but
shifting \emph{when} the cheap model is trusted, consistent with its
cost-side rather than accuracy-side value observed during
calibration~(\S\ref{sec:calibration}).

\FloatBarrier

% ---------------------------------------------------------------------------
% 7.3  Onboarding
% ---------------------------------------------------------------------------
\subsection{Onboarding a New Fixer}
\label{sec:results-onboarding}

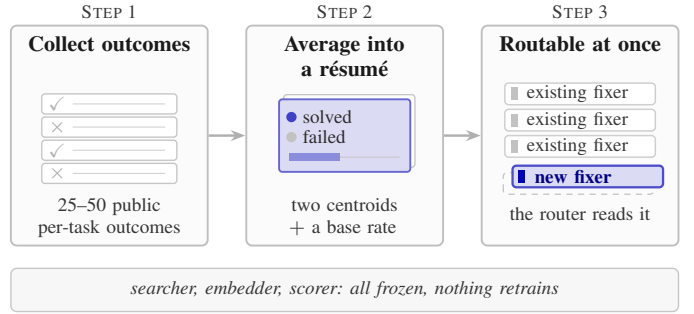
\begin{figure}[!tbp]
\centering
% ===========================================================================
% fig-onboarding.tex — onboarding a new fixer in three steps.
% BODY ONLY (tikzpicture in a \resizebox). Wrapper/caption/label live in the
% section file; mounts as a single-column \begin{figure}[!tbp].
% Requires \input{figures/fig-style} in the preamble.
%
% PROVENANCE
%   Schematic, illustrative only, no measured quantities. The only quantity
%   drawn is the documented spec constant 25--50 public per-task outcomes,
%   which is stated with its source in \S\ref{sec:results-onboarding} and
%   Table~\ref{tab:router-recipe}; nothing else in the figure is a number,
%   and no model or benchmark is named.
%
%   Glyph continuity: the r\'esum\'e card in step 2 and the fixer chips plus
%   dashed open slot in step 3 repeat the glyphs used in fig-pipeline.tex
%   (the r\'esum\'e shelf beneath the router, and the fixer-pool chips), so a
%   reader recognises the same objects across the two figures.
%
%   Colour contract (fig-style.tex): cSys tints only what onboarding adds
%   (the new r\'esum\'e, the new chip); everything already in the system is
%   neutral gray, which is also the point of the frozen strip along the
%   bottom. Fills differ in lightness, so the figure survives grayscale.
% ===========================================================================
\resizebox{\linewidth}{!}{%
\begin{tikzpicture}[
  x=1cm, y=1cm,
  panel/.style={rounded corners=2.2pt, line width=0.7pt, draw=gray!50,
                fill=gray!3, anchor=center},
  kick/.style={font=\scriptsize\scshape, text=gray!45!black, anchor=south,
               inner sep=0pt},
  head/.style={font=\footnotesize\bfseries, text=gray!20!black,
               anchor=north, align=center, inner sep=0pt},
  micro/.style={font=\scriptsize, text=gray!42!black, anchor=north,
                align=center, inner sep=0pt},
  flow/.style={-{Stealth[length=1.9mm,width=1.4mm]}, gray!62,
               line width=0.9pt},
]

% =========================================================================
% Geometry: three 2.55cm panels, 2.85cm tall, centred on y=0, with 0.50cm
% gaps for the arrows.  Centres: 1.275 / 4.325 / 7.375; drawing width 8.65.
% =========================================================================
\def\Pw{2.55} \def\Ph{2.85}

\node[panel, minimum width=\Pw cm, minimum height=\Ph cm] (s1) at (1.275,0) {};
\node[panel, minimum width=\Pw cm, minimum height=\Ph cm] (s2) at (4.325,0) {};
\node[panel, minimum width=\Pw cm, minimum height=\Ph cm] (s3) at (7.375,0) {};

\node[kick] at (1.275,1.52) {Step 1};
\node[kick] at (4.325,1.52) {Step 2};
\node[kick] at (7.375,1.52) {Step 3};

\draw[flow] (s1) -- (s2);
\draw[flow] (s2) -- (s3);

% =========================================================================
% Shared per-panel layout: heading north-anchored at y = 1.30, glyph in the
% band y = -0.70 .. 0.70, one-line-or-two caption north-anchored at -0.80.
% =========================================================================

% ---- Step 1 — collect a short public outcome record ---------------------
\node[head, text width=2.35cm] at (1.275,1.30) {Collect outcomes};
\foreach \i/\m/\c in {0/{$\times$}/{gray!68}, 1/{$\checkmark$}/{gray!58},
                      2/{$\times$}/{gray!68}, 3/{$\checkmark$}/{gray!58}}{
  \draw[draw=gray!45, fill=white, line width=0.5pt, rounded corners=1.2pt]
    (0.400,-0.62+0.30*\i) rectangle ++(1.75,0.26);
  \node[font=\scriptsize, text=\c, inner sep=0pt]
    at (0.60,-0.49+0.30*\i) {\m};
  \draw[gray!35, line width=0.4pt]
    (0.80,-0.49+0.30*\i) -- (2.00,-0.49+0.30*\i);
}
\node[micro, text width=2.35cm] at (1.275,-0.80)
  {25--50 public\\per-task outcomes};

% ---- Step 2 — average into a résumé (glyph mirrors fig-pipeline.tex) ----
\node[head, text width=2.35cm] at (4.325,1.30) {Average into\\a r\'esum\'e};
\draw[gray!45, fill=white, line width=0.5pt, rounded corners=1.2pt]
  (3.535,-0.41) rectangle ++(1.70,0.94);
\draw[cSys!60, fill=cSys!14, line width=0.7pt, rounded corners=1.2pt]
  (3.475,-0.47) rectangle ++(1.70,0.94);
\fill[cSys!75] (3.635,0.24)  circle (1.9pt);
\fill[gray!48] (3.635,-0.02) circle (1.9pt);
\node[font=\scriptsize, text=gray!35!black, anchor=west, inner sep=0pt]
  at (3.775,0.245) {solved};
\node[font=\scriptsize, text=gray!35!black, anchor=west, inner sep=0pt]
  at (3.775,-0.015) {failed};
\draw[gray!45, line width=0.4pt] (3.615,-0.28) -- (5.035,-0.28);
\fill[cSys!45] (3.615,-0.33) rectangle (4.265,-0.23);
\node[micro, text width=2.4cm] at (4.325,-0.80)
  {two centroids\\$+$ a base rate};

% ---- Step 3 — routable at once (chips mirror fig-pipeline.tex) ----------
\node[head, text width=2.35cm] at (7.375,1.30) {Routable at once};
\foreach \i in {0,1,2}{
  \draw[draw=gray!38, fill=white, line width=0.55pt, rounded corners=1.4pt]
    (6.400,0.40-0.34*\i) rectangle ++(1.95,0.28);
  \fill[gray!48] (6.480,0.46-0.34*\i) rectangle ++(0.08,0.16);
  \node[font=\scriptsize, text=gray!45!black, anchor=west, inner sep=0pt]
    at (6.680,0.54-0.34*\i) {existing fixer};
}
% the open slot, and the new fixer landing in it
\draw[draw=gray!45, dash pattern=on 0.8mm off 0.7mm, line width=0.5pt,
      rounded corners=1.4pt] (6.380,-0.76) rectangle ++(1.95,0.28);
\draw[draw=cSys!70, fill=cSys!14, line width=0.85pt, rounded corners=1.4pt]
  (6.500,-0.67) rectangle ++(1.95,0.28);
\fill[cSys] (6.580,-0.61) rectangle ++(0.08,0.16);
\node[font=\scriptsize\bfseries, text=cSys!72!black, anchor=west,
      inner sep=0pt] at (6.780,-0.53) {new fixer};
\node[micro, text width=2.45cm] at (7.375,-0.92)
  {the router reads it};

% =========================================================================
% The frozen strip: nothing retrains
% =========================================================================
\draw[gray!45, fill=gray!6, line width=0.5pt, rounded corners=2.2pt]
  (0,-2.28) rectangle (8.65,-1.72);
\node[font=\scriptsize\itshape, text=gray!42!black] at (4.325,-2.00)
  {searcher, embedder, scorer: all frozen, nothing retrains};

\end{tikzpicture}}
\caption{\textbf{Onboarding a fixer costs a r\'esum\'e, not a training
run.} A short public outcome record averages into two centroids and a base
rate, and the router reads that r\'esum\'e at inference time. No component
is retrained.}
\label{fig:onboarding}
\end{figure}

Adding a new fixer requires no retraining.
A fixer's \emph{r\'esum\'e} consists of two averaged embeddings (one
over its solved tasks, one over its failed tasks) plus a base solve
rate, all buildable from 25--50 public per-task outcomes
(Table~\ref{tab:router-recipe}).
The embedder is frozen infrastructure and \modelname never retrains; the
router simply reads each new r\'esum\'e at inference time
(Figure~\ref{fig:onboarding}).
This design keeps the pool extensible: a fixer released tomorrow can be
routed to as soon as a short public-benchmark run produces its outcome
record.
Appendix~\ref{apx:bench-fresh} bounds what such public outcomes can
carry: on our fresh calibration tasks every fixer's public rate drops
and the public ordering does not survive.
A r\'esum\'e built from public outcomes is therefore a cold-start
device, adequate for admitting a new fixer to the pool but not for
fine-grained ranking, and the thresholds used in this paper were
calibrated on freshly measured labels for exactly this reason.

% Section 8, Calibration.
\section{Calibration on Fresh Tasks}
\label{sec:calibration}

\subsection{The Label Run}
\label{sec:calib-labels}

\sysname's router assigns each incoming task to a fixer on the basis of per-task r\'{e}sum\'{e}s and cost thresholds (\S\ref{sec:system}).
Both mechanisms require outcome labels: which fixer solved which task, and at what price.
Public leaderboard data supplied these labels for several languages but contained zero usable Python outcomes for any fixer in the pool.
Benchmark contamination policies lock those results away, and Appendix~\ref{apx:bench-fresh} documents why leaderboard-derived labels mislead when applied outside their original distribution.
We therefore collected fresh labels: 100 Python bugs drawn from 2026 repositories, each attempted by all four fixers twice, once cold and once with \modelname's handoff injected (99 of the 100 yielded a valid paired comparison).
This paired design let every task serve as its own control, isolating the handoff's effect from task difficulty.

The same 100 tasks exposed a problem with \modelname's reproduction claims.
Replaying those claims against ground truth revealed that 32\% were genuine and 56\% demonstrably false.
The verify-then-strip gate described in \S\ref{sec:system} was built as a direct consequence, before any benchmark contact.

\subsection{The Handoff Redistributes}
\label{sec:calib-redist}

Figure~\ref{fig:redistribution} shows the paired comparison; exact rates and intervals appear in Appendix~\ref{apx:redistribution}.
Confidence intervals here and in the appendix are 95\% percentile
intervals from a bootstrap resampling tasks~\citep{bootstrap}.
The handoff lifted the three weaker fixers: \fxopus rose from 48.5\% to 53.5\% ($+5.1$\,pp, $p=0.125$), \fxkimi from 52.5\% to 56.6\% ($+4.0$\,pp, $p=0.481$), and \fxflash from 55.6\% to 57.6\% ($+2.0$\,pp,
$p=0.774$).
The strongest fixer, \fxgpt, moved in the opposite direction, dropping from 60.6\% to 56.6\% ($-4.0$\,pp, $p=0.424$).
Pooled across all four, the effect is $+1.8$ percentage points with a 95\% confidence interval of $[-1.0, {+}4.5]$ and $p=0.401$, not statistically significant.
At $N{=}99$ every per-fixer test is underpowered; the confidence intervals, not the point estimates, carry the information.

What the pattern does establish is that fixer rankings survive handoff conditioning, with only one adjacent flip falling within the confidence interval.
That stability is what the router's r\'{e}sum\'{e} assumption requires.
The pattern is consistent with redistribution rather than addition: solving ability shifts toward the cheaper models, which is exactly the shape cost-aware routing needs, though no single contrast reaches significance.
The redistribution pattern recurs on the benchmark's router-selected subset, though that comparison is uncontrolled and serves only as corroboration; exact numbers appear in Appendix~\ref{apx:pro-handoff}.

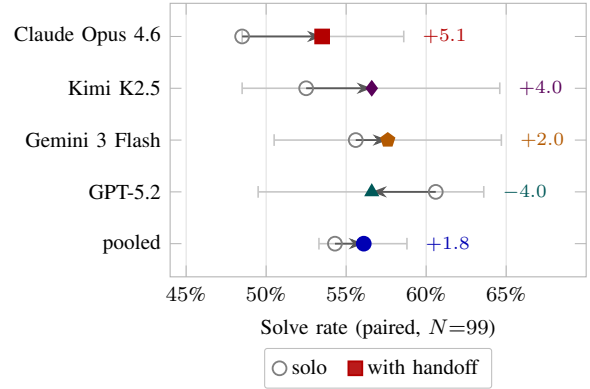
\begin{figure}[!tbp]
  \centering
  \begin{tikzpicture}
\begin{axis}[
  paperaxis,
  width=0.80\linewidth,
  height=5.2cm,
  xmin=44, xmax=70,
  xtick={45,50,55,60,65},
  xticklabel={\pgfmathprintnumber{\tick}\%},
  xlabel={Solve rate (paired, $N{=}99$)},
  symbolic y coords={{claude},{kimi},{flash},{gpt},{pooled}},
  ytick=data,
  % internal symbolic-coordinate keys above stay bare identifiers for
  % pgfplots bookkeeping; the RENDERED tick labels go through the fixer
  % macros so no bare family name ever reaches the page.
  yticklabels={\fxopus,\fxkimi,\fxflash,\fxgpt,pooled},
  y dir=reverse,
  enlarge y limits=0.16,
  xmajorgrids=true,
  legend style={at={(0.5,-0.26)}, anchor=north, legend columns=2,
                /tikz/every even column/.append style={column sep=6pt}},
  legend cell align=left,
]

% ---- paired-delta 95% CI, anchored at the solo rate --------------------
% [solo + CI_low, solo + CI_high]; see the provenance block above.
\draw[gray!45, line width=0.6pt] (axis cs:48.5,{claude}) -- (axis cs:58.6,{claude});
\draw[gray!45, line width=0.6pt] (axis cs:48.5,{kimi})   -- (axis cs:64.6,{kimi});
\draw[gray!45, line width=0.6pt] (axis cs:50.5,{flash})  -- (axis cs:64.7,{flash});
\draw[gray!45, line width=0.6pt] (axis cs:49.5,{gpt})    -- (axis cs:63.6,{gpt});
\draw[gray!45, line width=0.6pt] (axis cs:53.3,{pooled}) -- (axis cs:58.8,{pooled});
% end caps
\draw[gray!45, line width=0.6pt] ([yshift=2.2pt] axis cs:48.5,{claude}) -- ([yshift=-2.2pt] axis cs:48.5,{claude});
\draw[gray!45, line width=0.6pt] ([yshift=2.2pt] axis cs:58.6,{claude}) -- ([yshift=-2.2pt] axis cs:58.6,{claude});
\draw[gray!45, line width=0.6pt] ([yshift=2.2pt] axis cs:48.5,{kimi})   -- ([yshift=-2.2pt] axis cs:48.5,{kimi});
\draw[gray!45, line width=0.6pt] ([yshift=2.2pt] axis cs:64.6,{kimi})   -- ([yshift=-2.2pt] axis cs:64.6,{kimi});
\draw[gray!45, line width=0.6pt] ([yshift=2.2pt] axis cs:50.5,{flash})  -- ([yshift=-2.2pt] axis cs:50.5,{flash});
\draw[gray!45, line width=0.6pt] ([yshift=2.2pt] axis cs:64.7,{flash})  -- ([yshift=-2.2pt] axis cs:64.7,{flash});
\draw[gray!45, line width=0.6pt] ([yshift=2.2pt] axis cs:49.5,{gpt})    -- ([yshift=-2.2pt] axis cs:49.5,{gpt});
\draw[gray!45, line width=0.6pt] ([yshift=2.2pt] axis cs:63.6,{gpt})    -- ([yshift=-2.2pt] axis cs:63.6,{gpt});
\draw[gray!45, line width=0.6pt] ([yshift=2.2pt] axis cs:53.3,{pooled}) -- ([yshift=-2.2pt] axis cs:53.3,{pooled});
\draw[gray!45, line width=0.6pt] ([yshift=2.2pt] axis cs:58.8,{pooled}) -- ([yshift=-2.2pt] axis cs:58.8,{pooled});

% ---- solo -> handoff connectors ---------------------------------------
\draw[-{Stealth[length=2.0mm]}, black!65, line width=0.7pt] (axis cs:48.5,{claude}) -- (axis cs:53.5,{claude});
\draw[-{Stealth[length=2.0mm]}, black!65, line width=0.7pt] (axis cs:52.5,{kimi})   -- (axis cs:56.6,{kimi});
\draw[-{Stealth[length=2.0mm]}, black!65, line width=0.7pt] (axis cs:55.6,{flash})  -- (axis cs:57.6,{flash});
\draw[-{Stealth[length=2.0mm]}, black!65, line width=0.7pt] (axis cs:60.6,{gpt})    -- (axis cs:56.6,{gpt});
\draw[-{Stealth[length=2.0mm]}, black!65, line width=0.7pt] (axis cs:54.3,{pooled}) -- (axis cs:56.1,{pooled});

% ---- solo (open, gray) -------------------------------------------------
\addplot[only marks, papermark, color=cBase, mark=o, mark options={fill=white}]
  coordinates {(48.5,{claude}) (52.5,{kimi}) (55.6,{flash})
               (60.6,{gpt}) (54.3,{pooled})};
\addlegendentry{solo}

% ---- with handoff (filled, per-fixer colour + distinct marker) ---------
\addplot[only marks, papermark, color=cOpus, mark=square*]
  coordinates {(53.5,{claude})};
\addlegendentry{with handoff}
\addplot[only marks, papermark, color=cKimi, mark=diamond*, forget plot]
  coordinates {(56.6,{kimi})};
\addplot[only marks, papermark, color=cFlash, mark=pentagon*, forget plot]
  coordinates {(57.6,{flash})};
\addplot[only marks, papermark, color=cGpt, mark=triangle*, forget plot]
  coordinates {(56.6,{gpt})};
\addplot[only marks, papermark, color=cSys, mark=*, forget plot]
  coordinates {(56.1,{pooled})};

% ---- delta labels ------------------------------------------------------
\node[font=\scriptsize, text=cOpus,  anchor=west] at (axis cs:59.2,{claude}) {$+5.1$};
\node[font=\scriptsize, text=cKimi,  anchor=west] at (axis cs:65.2,{kimi})   {$+4.0$};
\node[font=\scriptsize, text=cFlash, anchor=west] at (axis cs:65.3,{flash})  {$+2.0$};
\node[font=\scriptsize, text=cGpt,   anchor=west] at (axis cs:64.2,{gpt})    {$-4.0$};
\node[font=\scriptsize, text=cSys,   anchor=west] at (axis cs:59.4,{pooled}) {$+1.8$};

\end{axis}
\end{tikzpicture}
  \caption{\textbf{The handoff pattern is redistributive, not additive.}
  Paired comparison ($N{=}99$ tasks, four fixers).  Open circles: solo
  rates; filled markers: with-handoff.  Horizontal bars span the 95\%
  CI of each paired delta, anchored at the solo rate.  The three weaker
  fixers gain; the strongest loses.  All deltas are directional only
  ($p>0.05$ at this sample size).}
  \label{fig:redistribution}
\end{figure}

\subsection{The Router Design Space}
\label{sec:calib-design}

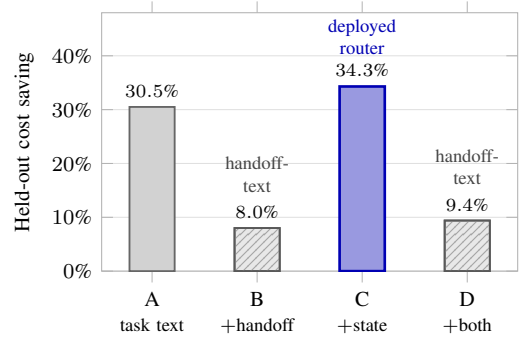
\begin{figure}[!tbp]
  \centering
  % ===========================================================================
% fig-design-space.tex — router input design space: held-out cost saving by
% feature variant.
% BODY ONLY (tikzpicture). Wrapper/caption/label live in the section file.
% Requires \input{figures/fig-style} in the preamble.
%
% PROVENANCE — every number.
%   Primary source: internal records, the definitive 4-row comparison on the
%   combined deployable router, column "saving %" (all four rows at the SAME matched accuracy .606,
%   fold-held-out, N=99, anchor = always-best gpt at $0.406 / .606):
%     A  task text only                              +30.5%  ($0.2824/task)
%     B  task text + handoff text                    + 8.0%  ($0.3734/task)
%     C  task text + hidden states                   +34.3%  ($0.2668/task)
%     D  task text + handoff text + hidden states    + 9.4%  ($0.3677/task)
%   Cross-checked against the same source's head-ablation table (LR rows:
%   30.5 / 8.0 / 34.3 / 9.4 — identical). Variant C, the text+hidden-state
%   blend, is the router's head; variant A is the text-only comparison.
%   The "handoff-text features hurt cost routing" reading is the source's
%   own: "The handoff-text embedding is a bad blend partner: averaging it in
%   (variants B and D) drags the saving down to ~8-9% ... so more features is
%   not better, the right feature is."
%   Fold stability (context, not plotted): A 3/5, B 1/5, C 3/5, D 4/5 vs
%   anchor; variant C's ~4pp gain over A is explicitly "not bankable at
%   N=99".
% ===========================================================================
\begin{tikzpicture}
\begin{axis}[
  paperaxis,
  width=0.80\linewidth,
  height=5.0cm,
  ybar,
  bar width=17pt,
  ymin=0, ymax=48,
  ytick={0,10,20,30,40},
  yticklabel={\pgfmathprintnumber{\tick}\%},
  ylabel={Held-out cost saving},
  symbolic x coords={{A},{B},{C},{D}},
  xtick={{A},{B},{C},{D}},
  xticklabels={A\\{\scriptsize task text}, B\\{\scriptsize $+$handoff},
               C\\{\scriptsize $+$state}, D\\{\scriptsize $+$both}},
  xticklabel style={font=\footnotesize, align=center},
  enlarge x limits=0.16,
  ymajorgrids=true,
  nodes near coords={\pgfmathprintnumber[fixed, fixed zerofill, precision=1]{\pgfplotspointmeta}\%},
  nodes near coords style={font=\scriptsize},
]

% ---- A: text-only baseline --------------------------------------------
\addplot[fill=cBase!35, draw=cBase!80!black, bar shift=0pt]
  coordinates {({A},30.5)};

% ---- B and D: variants carrying handoff-text features (hatched) --------
\addplot[fill=cBase!18, draw=cBase!70!black, bar shift=0pt,
         postaction={pattern=north east lines, pattern color=cBase!70}]
  coordinates {({B},8.0) ({D},9.4)};

% ---- C: the winner, the deployed router --------------------------------
\addplot[fill=cSys!40, draw=cSys, line width=1.0pt, bar shift=0pt]
  coordinates {({C},34.3)};

% ---- winner callout ----------------------------------------------------
\node[font=\scriptsize, text=cSys, anchor=south, align=center]
  at (axis cs:{C},38.6) {deployed\\router};

% ---- the two handoff-text variants -------------------------------------
\node[font=\scriptsize, text=gray!55!black, anchor=south, align=center]
  at (axis cs:{B},13.0) {handoff-\\text};
\node[font=\scriptsize, text=gray!55!black, anchor=south, align=center]
  at (axis cs:{D},14.4) {handoff-\\text};

\end{axis}
\end{tikzpicture}
  \caption{\textbf{Which router features buy cost savings.}
  Held-out cost saving against an always-best-model anchor, all variants
  pinned to the same solve rate ($.606$).  Adding \modelname's hidden
  states (variant~C, the routing blend) lifts savings from $30.5\%$ to $34.3\%$.
  Variants with handoff-text features (B,~D) collapse to $8$--$9\%$.
  The C-vs-A gap is fold-fragile at $N{=}99$.}
  \label{fig:design-space}
\end{figure}

We compared four input-feature variants under an offline simulation of the capped cost protocol, pinning each to the same matched solve rate so that the bars in Figure~\ref{fig:design-space} differ only in spend.
Variant~A, using task text alone, saves 30.5\% on total cost per solve at a matched solve rate.
Adding \modelname's hidden states produces variant~C, the routing blend, raising the saving to 34.3\%.
The two variants that incorporate handoff-text features tell a sharply different story: variant~B collapses to 8.0\% and variant~D to 9.4\%.
The lesson is concrete.
The searcher's internal state helps the router, while the handoff's text helps the fixers; feeding the handoff memo to the router actively hurts.

Supporting signals reinforce this split.
As an outcome predictor, the hidden state reaches an AUC of 0.600 (fold-stable in four of five folds), while handoff text manages only 0.510.
A simple logistic regression beat a multi-layer perceptron in every comparison, so the simpler scorer shipped.
The full grid appears in Appendix~\ref{apx:design-space}.
The C-versus-A cost gap, a different quantity from the AUC separation,
holds in only three of five folds, and the gate-level caveat of \S\ref{sec:results} applies: at deployment the hidden state's informational contribution cannot be fully separated from its threshold-shifting effect.
On the calibration labels, the cost-aware router matched the strongest fixer's solve rate at 29\% lower total cost per solve (\$0.49 versus \$0.69).

Calibration settled four design choices: the routing blend (variant~C), the matched-point threshold ($\theta{=}0.30$), logistic regression as the scoring function, and blanket handoff injection.
It left one question open.
\fxgpt was the strongest fixer on these fresh tasks, a ranking the benchmark later inverted (\S\ref{sec:results}).

\FloatBarrier

% Section 9, Limitations.
\section{Limitations}
\label{sec:limitations}

Our headline result rests on a single benchmark's Python slice,
comprising 266 tasks.  While the searcher's localization ability
transfers across languages (\S\ref{sec:training}), that evidence covers
the searcher alone, not the full routed system.  Cross-benchmark
generality remains untested, and the ablation finding, that blanket
assignment to the cheapest fixer ties the routed system, is a property
of this task pool; it need not transfer to other pools or benchmarks.
The 266 tasks themselves span three repositories, with per-repository
recall ranging from 0.472 to 0.671
(Appendix~\ref{apx:localization}); the effective diversity is
closer to three codebases than to 266 independent draws.

At $n{=}266$ the headline is a one-task edge, which is precisely why we
claim a \emph{match} and nothing stronger.  The robustness evidence is
structural: both the routed system and the no-router ablation sit
$+3.60$ percentage points above the blind-mixing line, so on this pool
the margin over random traffic splitting is carried by the handoff
rather than by the routing decision.  Like the ablation finding itself,
this attribution is specific to this pool and benchmark and can change
with either.

The calibration label run uses $n{=}100$ tasks.  No handoff contrast in
this paper reaches statistical significance, on the calibration labels
or on the benchmark subset; the mechanism rests on the consistency of
the pattern across two independent settings, not on any single test.

The benchmark-side handoff comparisons themselves come from
router-selected subsets rather than random assignment, so they carry
observational caveats.  The controlled reference point is the paired
calibration study on held-out tasks (\S\ref{sec:calibration}).

Contamination control is one-sided.  The searcher's training corpus
excludes a 23-repository blocklist covering every SWE-bench Pro and
SWE-bench Verified repository, enforced by a programmatic gate on the
frozen training file.  No such control is possible on the fixer side:
the three evaluation repositories are public, we have no visibility
into the four fixers' training data, and differential fixer-side
contamination would distort the relative solve rates that the
r\'esum\'es and the calibration are fit to.

We also did not run a pass-through arm in which fixers receive the
handoff with its unverified reproduction claims intact, so the guard's
contribution to the headline solve rate is asserted from the claim
census (\S\ref{sec:results}) rather than measured; a pass-through
ablation is future work.

All costs are provider list prices recorded at measurement time.
Prices drift, and one fixer was served as a quantized build through a
pinned provider, so cost conclusions are snapshots rather than stable
constants.  The cost-accounting asymmetry described in
\S\ref{sec:setup} (all-in for the system versus API-only for solo
baselines) is conservative in our disfavor, but it is still an
accounting choice that readers should weigh.

\modelname achieves an overall localization recall of 0.566, a moderate
figure drawn from a single sampled inference pass whose variance is
uncharacterized.  The system-level result shows that this recall
suffices for effective routing; it does not establish \modelname as a
state-of-the-art localizer.

All absolute solve rates reported here are specific to the capped,
silent-budget, auto-submit evaluation tier defined in
\S\ref{sec:setup}.  Under other tiers these numbers would change.  Our
protocol comparisons (\S\ref{sec:results}) show that such tier choices
are worth percentage points of solve rate, underscoring the relativity
of any single absolute figure.

Finally, the hidden-state features that drive routing are tied to the
exact \modelname checkpoint.  Retraining the searcher would require
re-extracting hidden states and refitting the router heads.  The
zero-retrain onboarding property applies to adding new fixers, not to
changing the searcher itself.

% Section 10, Conclusion.

\section{Conclusion}
\label{sec:conclusion}

\sysname shows that engaging with the repository before dispatch pays, though not where we expected: the verified handoff, not the routing decision, carries the result.
A small trained searcher, \modelname, explores the repository and writes a structured handoff whose reproduction claims are verified in a sandbox (false claims are stripped), and whose hidden states feed a r\'{e}sum\'{e} router that selects the downstream fixer.
On the benchmark's full Python slice, this pipeline solves 159 of 266 problems, matching the best single frontier model (158 of 266) at about a fifth of the total cost per solve, with the searcher's own compute contributing a rounding error to the budget.

Two mechanism-level findings underpin this result.
The handoff appears to redistribute rather than add ability, lifting the three cheaper fixers while slightly hurting the strongest, a directional pattern at $N{=}99$.
The searcher's hidden states, meanwhile, change which tasks the router trusts to the cheap model, yet feeding the handoff's text directly to the router hurts it.
Because adding a new fixer requires no retraining, the system is built for a model market that changes quarter by quarter.

Natural next steps include wider benchmark coverage across languages and problem domains.
Equally valuable would be measuring how quickly a newly released fixer can be onboarded into the live pipeline without retraining the searcher or the router.

% References start on a fresh page; \clearpage also flushes any pending
% floats so nothing from the main body or appendix interleaves with the
% bibliography. \balance evens out the two columns of the final body page.
\balance
\clearpage
\bibliographystyle{IEEEtranN}
\bibliography{references}

% Appendix likewise starts on its own page, so its floats can only land on
% appendix pages.
\clearpage
% \balance swaps the output routine globally and stays in force for every
% later page, so the one issued before the references was still balancing
% the appendix. Turn it off here: with pinned floats an unbreakable table
% cannot be split, so balancing pushed whole tables into the right column
% and left the left one nearly empty.
\nobalance
% Appendix floats are pinned with [H], so leftover vertical space must be
% allowed to collect at column bottoms rather than being stretched between
% sections. Applies from here to the end of the document only.
\raggedbottom
% Appendix typography one step down (10pt -> 9pt body, footnotesize captions).
% Standard venue practice for supplementary material; scoped to the appendix
% only because it is issued after the references \clearpage. Float bodies are
% shrunk a further step with \footnotesize inside each wrapper in
% sections/99_appendix.tex.
\small
\captionsetup{font=footnotesize}
\appendices
% Appendix, sections A through N.

% ============================================================================
% A — The Prior Routing Audit
% ============================================================================
\section{The Prior Routing Audit}
\label{apx:audit}

The cost-routing premise of \S\ref{sec:problem} rests on a zero-cost replay of
published per-task outcome matrices.  Table~\ref{tab:apx-audit} records the
structural statistics behind that replay: set containment, the unique-solver
shell, and the gap between the best learned router and always picking the
strongest model, across three benchmarks and matched four-model pools.

\begin{table}[H]
\centering
\footnotesize
\caption{\textbf{Why we do not route for accuracy.}
Zero-cost audit of published per-task outcome matrices on matched four-model
pools.  \emph{Containment} measures solve-set nesting; the
\emph{unique-solver shell} is the share of routable tasks only one model
solves.  On all three benchmarks, no learned router beat always calling the
strongest model.  This is background motivation, not a system result.}
\label{tab:apx-audit}
\begin{tabular}{@{}lrrr@{}}
\toprule
 & Verified & Multiling. & Pro \\
\midrule
Best single model      & 0.768 & 0.726 & 0.449 \\
Mean containment       & 0.941 & 0.912 & 0.773 \\
Unique-solver shell    & 21.5\% & 20.8\% & 30.8\% \\
\addlinespace
Best router $-$ best model & $-0.006$ & $+0.010$ & $-0.004$ \\
\quad $p$              & 0.66 & 0.24 & 0.90 \\
\bottomrule
\end{tabular}

\end{table}

% ============================================================================
% B — Router Threshold Sweep
% ============================================================================
\section{Router Threshold Sweep}
\label{apx:theta-sweep}

Table~\ref{tab:apx-theta-sweep} sweeps the routing threshold $\theta$ that
the calibration's matched-point rule selects.  At $\theta{=}0.25$ every task
clears the cheapest fixer's gate and the router degenerates into the no-router
ablation: 159 solves at \$0.227 per solve, identical to always calling
\fxkimi.  Raising the threshold to the operating point $\theta{=}0.30$ diverts
three tasks to \fxflash at a marginal cost of \$0.003 per solve, without
changing the solve count.  Above the operating point, the router's growing
caution pushes progressively more tasks onto expensive fixers: cost per task
rises monotonically while the solve count ceases to be exactly measurable,
because the diverted tasks lack a with-handoff outcome on their new fixer.
We report those rows as bracketing intervals rather than point estimates (the
table note details the convention).  On this benchmark, routing collapses to
a cost-allocation decision: no threshold examined buys additional solves
beyond the 159 that the cheapest fixer already delivers.

\begin{table}[H]
\centering
\footnotesize
\caption{\textbf{The gate threshold sweep.}
Raising $\theta$ pushes tasks off the cheapest fixer onto more expensive ones:
cost per task rises monotonically while the solve count ceases to be exactly
measurable, since diverted tasks lack a with-handoff outcome on their new
fixer.  The operating point $\theta{=}0.30$ routes all but three tasks to
\fxkimi and is the last threshold at which every number is exact.}
\label{tab:apx-theta-sweep}
% ===========================================================================
% tab-apx-theta-sweep.tex — APPENDIX. Gate threshold sweep: what the router
% does to routing, solves and cost as theta moves.
% BODY ONLY (tabular). Wrapper/caption/label live in the appendix file; the
% appendix wrapper supplies \footnotesize, so this body sets no font size,
% only a tighter \tabcolsep (eight columns in one 3.5in text column).
% Provenance: code/router/router_spec.json, data/receipts/outcomes_per_task.json.
%
% Convention: exact rows (0.25, 0.30) print a point solve count; bounded rows
%   print the bracketing interval [lo, hi] and NEVER a bare midpoint, here or
%   in caption or prose. The star marks a $/solve computed from a solo-filled
%   point estimate of the solve count; $/task carries no star because it is
%   fixed by the routing split. theta = 0.30 is the operating point. Above the
%   operating point, tasks passing no gate go to the most capable fixer
%   (stated in the note); at 0.25/0.30 the fall-through never fires.
% ===========================================================================
\begingroup\setlength{\tabcolsep}{2pt}%
\begin{tabular}{@{}lrrrrrrr@{}}
\toprule
& \multicolumn{4}{c}{Routing split (tasks)} & & & \\
\cmidrule(lr){2-5}
$\theta$
  & \parbox[b]{0.42in}{\raggedleft\fxkimi}
  & \parbox[b]{0.42in}{\raggedleft\fxflash}
  & \parbox[b]{0.42in}{\raggedleft\fxgpt}
  & \parbox[b]{0.42in}{\raggedleft\fxopus}
  & Solves & \$/task & \$/solve \\
\midrule
$0.25$              & 266 &  0 &  0 &  0 & 159        & \$0.136 & \$0.227 \\
$0.30^{\dagger}$    & 263 &  3 &  0 &  0 & 159        & \$0.137 & \$0.230 \\
$0.35$              & 236 & 19 & 10 &  1 & [149, 169] & \$0.154 & \$0.272$^{*}$ \\
$0.40$              & 203 & 34 & 28 &  1 & [132, 179] & \$0.178 & \$0.337$^{*}$ \\
$0.50$              & 124 & 34 & 72 & 36 & [94, 196]  & \$0.334 & \$0.672$^{*}$ \\
\bottomrule
\multicolumn{8}{@{}p{0.97\linewidth}@{}}{\footnotesize $^{\dagger}$Operating
point: the threshold used for every \sysname number in the paper.
$^{*}$\$/solve from a solo-filled point estimate of the solve count, since
tasks routed to a fixer with no measured with-handoff outcome take that
fixer's solo outcome; rows whose solve count is not measured exactly report
the bracketing interval instead, and no midpoint. Rows above the operating
point send tasks that pass no gate to the most capable fixer (\fxopus); at the
operating point every task passes at least one gate, so the fall-through never
fires. Solves are out of $n{=}266$.
All rows use the board's cost convention: measured fixer API spend plus a
pinned \$5.13 of searcher GPU and sandbox infrastructure, amortized over 266
tasks.}
\end{tabular}%
\endgroup

\end{table}

% ============================================================================
% C — Verify-then-Strip Outcome Classes
% ============================================================================
\section{Verify-then-Strip Outcome Classes}
\label{apx:verify-strip}

\S\ref{sec:results} reports the headline guard statistics; the full
failure-class census behind those numbers appears in
Table~\ref{tab:apx-verify-strip}, broken down by outcome category and by
handoff type.

\begin{table}[H]
\centering
\footnotesize
\caption{\textbf{What the verify-then-strip sandbox found in 266 handoffs.}
\emph{Top:} outcome class of each reproduction claim, replayed before any
fixer saw it ($50$ genuine, $174$ false and stripped).  \emph{Bottom:}
split by handoff type; spontaneous handoffs are more than twice as likely
to be genuine as forced ones (never pooled).  Forced-handoff percentages
use the $43$ of $60$ that claim a reproduction as their denominator.
The label-run row shows the same measurement on an earlier, easier task
set.}
\label{tab:apx-verify-strip}
\begin{tabular}{@{}lr@{}}
\toprule
Outcome of the replayed reproduction & $n$ \\
\midrule
Passed at base (claim stripped)     & 174 \\
Genuinely failed (claim kept)       & \phantom{0}50 \\
Claim not true                      & \phantom{0}16 \\
Import error                        & \phantom{0}12 \\
Non-zero exit, other                & \phantom{00}9 \\
Missing file                        & \phantom{00}3 \\
No reproduction block               & \phantom{00}1 \\
No command                          & \phantom{00}1 \\
\midrule
Total                               & 266 \\
\bottomrule
\end{tabular}

\vspace{4pt}

\begin{tabular}{@{}lrrr@{}}
\toprule
Handoff set & Claiming & Genuine & Stripped \\
\midrule
Spontaneous ($n{=}206$) & 206 & 22\% & 68\% \\
Forced ($n{=}60$)       & \phantom{0}43 & \phantom{0}9\% & 79\% \\
\midrule
All Pro handoffs ($n{=}266$) & 249 & 20\% & 70\% \\
Label-run baseline           & \phantom{0}91 & 32\% & 56\% \\
\bottomrule
\end{tabular}

\end{table}

% ============================================================================
% D — Localization by Repository
% ============================================================================
\section{Localization by Repository}
\label{apx:localization}

The per-task mean localization quality reported in Table~\ref{tab:localization}
pools three repositories that differ substantially in how many files their
gold patches touch.  Table~\ref{tab:apx-localization-repo} disaggregates by
repository and provides the solved-versus-unsolved cross-tab behind the
main-text numbers.

\begin{table}[H]
\centering
\footnotesize
\caption{\textbf{Localization by repository, and against outcome.}
\emph{Top:} \modelname's file-localization quality per repository; spread
tracks the number of gold files per task.  \emph{Bottom:} quality split
by whether the system solved the task.  Better localization correlates
weakly with solving ($+0.024$ recall), so localization is not a gate on
the outcome.}
\label{tab:apx-localization-repo}
\begin{tabular}{@{}lrrrr@{}}
\toprule
Repository & Recall & Prec. & All-gold & Gold files \\
\midrule
ansible ($n{=}96$)     & 0.472 & 0.846 & 10.4\% & 3.80 \\
openlibrary ($n{=}91$) & 0.574 & 0.766 & 28.6\% & 3.58 \\
qutebrowser ($n{=}79$) & 0.671 & 0.853 & 38.0\% & 2.84 \\
\midrule
All ($n{=}266$)        & 0.566 & 0.821 & 24.8\% & 3.44 \\
\bottomrule
\end{tabular}

\vspace{4pt}

\begin{tabular}{@{}lrrr@{}}
\toprule
Outcome & Recall & Prec. & All-gold \\
\midrule
Solved ($n{=}159$)   & 0.575 & 0.830 & 26.4\% \\
Unsolved ($n{=}107$) & 0.552 & 0.808 & 22.4\% \\
\bottomrule
\end{tabular}

\end{table}

% ============================================================================
% E — Capped-Tier Statistics
% ============================================================================
\section{Capped-Tier Statistics}
\label{apx:cap-stats}

The protocol findings in \S\ref{sec:results} quote selected cap-binding
rates and auto-submit rescue counts.  Table~\ref{tab:apx-cap-stats} gives
the complete picture for all three fixer arms.

\begin{table}[H]
\centering
\footnotesize
\caption{\textbf{How often the evaluation caps bound, and what
auto-submission was worth.}  All arms ran the official capped tier ($50$
turns, \$2.00 per attempt).  The turn cap is the binding constraint; the
cost cap bound only five times.  Auto-submission rescued $24$ of
\fxopus's solves and $33$ of \fxgpt's; without it, both anchors' rates
would fall roughly $9$--$13$ points.  The gold-patch control passes on
$265$ of $266$ tasks.}
\label{tab:apx-cap-stats}
\begin{tabular}{@{}lrrr@{}}
\toprule
 & \fxgpt & \fxopus & \fxkimi \\
\midrule
Hit 50-turn cap        & 36.5\% & 17.3\% & 48\% \\
Hit \$2 cost cap       & 0 & 5 & 0 \\
Max attempt cost       & \$1.59 & \$2.09 & --- \\
Auto-submitted at cap  & 94 & 48 & 126 \\
Solves rescued         & 33 & 24 & --- \\
Solves                 & 139 & 158 & 149 \\
Rate without rescue    & 39.8\% & 50.4\% & --- \\
\bottomrule
\end{tabular}

\end{table}

% ============================================================================
% F — Router Design Space, Full Grid
% ============================================================================
\section{Router Design Space, Full Grid}
\label{apx:design-space}

The calibration analysis in \S\ref{sec:calibration} summarizes the four
feature variants under the logistic-regression scorer.
Table~\ref{tab:apx-design-space} extends that comparison to include the MLP
head and provides the outcome-separation (AUC) measurements the cost numbers
rest on.

\begin{table}[H]
\centering
\footnotesize
\caption{\textbf{The complete router design space.}
\emph{Top:} held-out cost saving for every feature-set $\times$ scorer
combination, all pinned to solve rate $.606$.  Adding hidden states helps
under both scorers; handoff-text features collapse savings to
$8$--$9\%$.  The MLP halves every variant's saving.  At $N{=}99$ the
best variant is only $3/5$ fold-stable.  \emph{Bottom:} the hidden
state separates solved from failed tasks at AUC~$.600$, while
handoff-text embedding sits at chance.}
\label{tab:apx-design-space}
\begin{tabular}{@{}lrrrr@{}}
\toprule
Router features & \multicolumn{2}{c}{Saving} & Folds & Brier \\
\cmidrule(lr){2-3}
 & LR & MLP & & \\
\midrule
A\quad text only                   & 30.5\% & 15.4\% & 3/5 & 0.270 \\
B\quad $+$ handoff text            & \phantom{0}8.0\% & \phantom{0}5.3\% & 1/5 & 0.254 \\
C\quad $+$ hidden state \emph{(dep.)} & \textbf{34.3\%} & 18.7\% & 3/5 & \textbf{0.249} \\
D\quad $+$ both                    & \phantom{0}9.4\% & \phantom{0}4.2\% & 4/5 & 0.251 \\
\bottomrule
\end{tabular}

\vspace{4pt}

\begin{tabular}{@{}lrr@{}}
\toprule
Routing representation & Pooled AUC & Stable \\
\midrule
Hidden state (pre-decode, layer $-4$) & \textbf{0.600} & 4/5 \\
Task text                             & 0.561 & --- \\
Handoff text (prereg.\ baseline)      & 0.510 & --- \\
\bottomrule
\end{tabular}

\end{table}

% ============================================================================
% G — Redistribution, Exact Numbers
% ============================================================================
\section{Redistribution, Exact Numbers}
\label{apx:redistribution}

Figure~\ref{fig:redistribution} in the main text visualizes the paired
handoff ablation; Table~\ref{tab:apx-redistribution} gives the exact per-fixer
rates, deltas, confidence intervals, discordant-pair counts, and $p$-values
behind that figure.

\begin{table}[H]
\centering
\footnotesize
\caption{\textbf{The handoff pattern is redistributive, not additive.}
Exact per-fixer rates, deltas, and $p$-values behind
Figure~\ref{fig:redistribution} ($99$ paired tasks, $396$ attempts).
The three weaker fixers gain and the strongest loses; pooled effect is
$+1.8$\,pp with CI including zero.  No fixer reaches $p<0.05$; every
row is directional only.  $b$~counts rescued tasks, $c$~counts broken
ones.}
\label{tab:apx-redistribution}
\begin{tabular}{@{}lrrrrr@{}}
\toprule
Fixer & Solo & Handoff & $\Delta$ (95\% CI) & $b/c$ & $p$ \\
\midrule
\fxopus & 48.5 & 53.5 & $+5.1$ [$0.0$, $10.1$]  & 6/1  & 0.125 \\
\fxkimi   & 52.5 & 56.6 & $+4.0$ [$-4.0$, $12.1$] & 11/7 & 0.481 \\
\fxflash  & 55.6 & 57.6 & $+2.0$ [$-5.1$, $9.1$]  & 7/5  & 0.774 \\
\fxgpt    & 60.6 & 56.6 & $-4.0$ [$-11.1$, $3.0$] & 5/9  & 0.424 \\
\midrule
pooled & 54.3 & 56.1 & $+1.8$ [$-1.0$, $4.5$]  & 29/22 & 0.401 \\
\bottomrule
\end{tabular}

\end{table}

% ============================================================================
% H — Benchmark-Versus-Fresh Calibration Receipts
% ============================================================================
\section{Benchmark-Versus-Fresh Calibration Receipts}
\label{apx:bench-fresh}

The router's r\'esum\'es are built from public per-task outcomes, so the
router inherits whatever biases those outcomes carry.  Buying fresh labels was
not a design preference but a measured necessity.
Table~\ref{tab:apx-bench-fresh} records the calibration receipts that
motivated that decision: every fixer's public rate compared with its
freshly measured rate under a single controlled harness.

\begin{table}[H]
\centering
\footnotesize
\caption{\textbf{Why we bought fresh labels: public rates do not transfer
to unseen tasks.}  Each fixer's public multilingual rate versus its freshly
measured rate on $100$ post-cutoff Python tasks under one harness.
Every fixer drops; the ordering does not survive.  Rates use the full
$100$-task solo denominator; the paired analysis in
Appendix~\ref{apx:redistribution} uses the $99$ tasks with both arms.
These are calibration receipts, not a benchmark claim.}
\label{tab:apx-bench-fresh}
\begin{tabular}{@{}lrrrr@{}}
\toprule
Fixer & Public & Fresh (95\% CI) & Gap & Rank \\
\midrule
\fxflash & 0.727 & 0.550 [.450, .650] & $-17.7$ & \#1 $\rightarrow$ \#2 \\
\fxopus  & 0.720 & 0.480 [.380, .570] & $-24.0$ & \#1 $\rightarrow$ \#4 \\
\fxkimi  & 0.673 & 0.520 [.420, .610] & $-15.3$ & \#2 $\rightarrow$ \#3 \\
\fxgpt   & 0.667 & 0.610 [.510, .700] & $-5.7$ & \#3 $\rightarrow$ \#1 \\
\bottomrule
\end{tabular}

\end{table}

% ============================================================================
% I — Handoff Effect on Router-Selected Benchmark Subsets
% ============================================================================
\section{Handoff Effect on Router-Selected Benchmark Subset}
\label{apx:pro-handoff}

The redistribution pattern observed under controlled conditions
(\S\ref{sec:calib-redist}) also appears in the benchmark evaluation, where
the tasks the router sends to \fxkimi were available both with and without a
handoff.  Table~\ref{tab:apx-pro-handoff} reports that comparison.  The subset
was selected by the router rather than randomly assigned; it remains measured,
uncontrolled corroboration rather than a controlled experiment.

\begin{table}[H]
\centering
\footnotesize
\caption{\textbf{The handoff effect on the deployment path.}
The tasks routed to \fxkimi were also run without a handoff, giving a
same-task comparison: the workhorse gains $3.8$\,pp on $263$ tasks.
$b/c$ counts tasks solved only with the handoff versus only solo; the
exact McNemar test on those discordant pairs gives $p=0.245$.  The
router selected \emph{which} tasks enter the subset, not the
within-task contrast, so the pairing is intact, but the subset is not
randomly assigned and this table remains corroboration, not a
controlled measurement.  The three tasks routed elsewhere have no solo
arm for their fixer and are omitted.}
\label{tab:apx-pro-handoff}
% ===========================================================================
% tab-apx-pro-handoff.tex — APPENDIX. Deployment-path corroboration of the
% handoff effect, on the router-selected subset of the Pro evaluation.
% BODY ONLY (tabular). Wrapper/caption/label live in the appendix file.
%
% PROVENANCE — every cell. Derived from data/receipts/:
%   routing_scenarios.json `decisions` (theta = 0.30) gives the routed subset:
%     263 tasks -> kimi-k2-5, 3 -> gemini-3-flash.
%   outcomes_per_task.json:
%     with handoff = system_union_pairs[task]["kimi-k2-5"].resolved
%       -> 158/263 = 60.1%
%     solo, same tasks = solo[task]["kimi"].resolved
%       -> 148/263 = 56.3%
%     delta = +3.8pp
%     discordant pairs: handoff-only b=35, solo-only c=25 (both 123,
%       neither 80); exact two-sided McNemar on b+c=60 -> p = 0.245
%   Whole-benchmark anchors from the same file: Kimi K2.5 with handoff
%   159/266, solo 149/266.
%
%   WHY ONE ROW: the 3 tasks routed to Gemini 3 Flash have no solo Flash arm,
%   so they cannot form a paired subset; the caption's footnote says so.
%   The subset is router-selected, not randomly assigned, so the paired
%   label-run comparison stays the controlled citation and this table is
%   corroboration only; the McNemar p is printed with that caveat in the
%   caption.
% ===========================================================================
\begin{tabular}{@{}lrrrrrr@{}}
\toprule
Fixer & Routed $n$ & Handoff & Solo & $\Delta$ & $b/c$ & $p$ \\
\midrule
\fxkimi & 263 & 60.1\% & 56.3\% & $+3.8$ & 35/25 & 0.245 \\
\bottomrule
\end{tabular}

\end{table}

% ============================================================================
% J — Nine-Language Localization Detail
% ============================================================================
\section{Nine-Language Localization Detail}
\label{apx:languages}

Figure~\ref{fig:language-transfer} in the main text shows the per-language
$F_1$ bars; Table~\ref{tab:apx-languages} gives the full precision, recall,
$F_1$, and all-gold counts behind that figure, split by trained and
never-trained language pools.

\begin{table}[H]
\centering
\footnotesize
\caption{\textbf{Localization transfers to languages \modelname never saw
in training.}  Per-language precision, recall, and $F_1$ of spontaneous
handoffs against gold patch files.  The six never-trained languages
outperform the three trained ones ($F_1$~$0.630$ vs.\ $0.455$).
JavaScript and TypeScript are genuine soft spots; the TypeScript and C++
cells rest on $12$ assigned tasks each, with $7$ and $8$ spontaneous
handoffs analyzed ($\dagger$).  Multi-file recall is weak across
all languages ($0.26$--$0.34$).  Forced handoffs excluded; searcher
alone, one sampled draw.}
\label{tab:apx-languages}
\begin{tabular}{@{}lrrrrr@{}}
\toprule
Language & $n$ & P & R & $F_1$ & All-gold \\
\midrule
Go                  & 27 & 0.605 & 0.542 & 0.546 & 13/27 \\
JavaScript          & 26 & 0.500 & 0.365 & 0.389 & \phantom{0}7/26 \\
TypeScript$^{\dagger}$ & \phantom{0}7 & 0.333 & 0.357 & 0.343 & \phantom{0}2/7 \\
\addlinespace
\emph{trained pool} & 60 & 0.528 & 0.444 & 0.455 & 22/60 \\
\midrule
Java                & 27 & 0.722 & 0.574 & 0.612 & 11/27 \\
Ruby                & 36 & 0.833 & 0.661 & 0.701 & 18/36 \\
Rust                & 20 & 0.800 & 0.626 & 0.667 & 10/20 \\
PHP                 & 33 & 0.697 & 0.632 & 0.630 & 17/33 \\
C                   & 18 & 0.546 & 0.435 & 0.435 & \phantom{0}6/18 \\
C++$^{\dagger}$     & \phantom{0}8 & 0.688 & 0.813 & 0.708 & \phantom{0}6/8 \\
\addlinespace
\emph{never-trained} & 142 & 0.731 & 0.613 & \textbf{0.630} & 68/142 \\
\bottomrule
\end{tabular}

\end{table}

% ============================================================================
% K — Vault Decoding Comparison
% ============================================================================
\section{Vault Decoding Comparison}
\label{apx:vault}

The decoding finding reported in \S\ref{sec:decoding} is demonstrated on
a 100-task dial subset; Table~\ref{tab:apx-vault} gives the full A/B
comparison on the 450-task held-out vault that established the finding.

\begin{table}[H]
\centering
\footnotesize
\caption{\textbf{Sampled decoding is what makes the searcher commit.}
The same checkpoint on $450$ held-out issues, greedy versus temperature
$0.9$.  Find rate nearly triples, from higher commitment
($72\%$ vs.\ $21\%$) against an 18\% drop in per-handoff recall.  The
greedy arm lost $24$ episodes to infrastructure timeouts; the
$2.65\times$ figure adjusts for this.  The vault is excluded from all
training data.}
\label{tab:apx-vault}
\begin{tabular}{@{}lrr@{}}
\toprule
Metric ($n{=}450$) & Sampled & Greedy \\
\midrule
Find rate            & 0.306 & 0.110 \\
\quad 95\% CI        & [.273, .337] & --- \\
\quad timeout-matched ratio & \multicolumn{2}{r}{$2.65\times$ ($n{=}426$)} \\
\addlinespace
Emission rate        & 0.718 & 0.213 \\
Handoffs emitted     & 323 & \phantom{0}96 \\
\addlinespace
Recall per handoff   & 0.426 & 0.517 \\
Precision            & 0.643 & 0.726 \\
Files per handoff    & 1.93 & 2.05 \\
All-gold handoffs    & \phantom{0}71 & \phantom{0}27 \\
\addlinespace
Malformed            & \phantom{00}0 & \phantom{00}0 \\
\bottomrule
\end{tabular}

\end{table}

% ============================================================================
% L — Training Hyperparameters
% ============================================================================
\section{Training Hyperparameters}
\label{apx:hyperparams}

\S\ref{sec:training} describes the supervised recipe at the level
relevant to the contribution; Table~\ref{tab:apx-hyperparams} records the
full configuration as realized in the actual training run.

\begin{table}[H]
\centering
\footnotesize
\caption{\textbf{Supervised fine-tuning configuration, as run.}
Rank-$64$ LoRA on every linear projection, trained in \code{bf16} for
two epochs on packed $32$k-token blocks at $91.7\%$ fill; loss is on
assistant turns only.  All values are the realized run, not the planned
recipe.  The complete run cost \$86 of GPU time and peaked at
$24.2$\,GB.}
\label{tab:apx-hyperparams}
\begin{tabular}{@{}ll@{}}
\toprule
Setting & Value \\
\midrule
Base model        & Qwen2.5-Coder-7B-Instruct \\
Adaptation        & LoRA, $r{=}64$, $\alpha{=}128$ \\
Target modules    & all linear projections \\
                  & (no embedding, no output head) \\
Dropout           & 0 \\
Precision         & \texttt{bf16} \\
\midrule
Learning rate     & $1{\times}10^{-4}$, cosine to $10\%$ \\
Warmup            & 3\% of steps \\
Epochs            & 2 (3,182 steps) \\
Context           & 32k, packed blocks \\
Micro-batch       & $1{\times}32$k \\
Gradient accum.   & 8 \\
Gradient ckpt.    & on \\
Seed              & 3407 \\
\midrule
Tokens per epoch  & 382.4M (91.7\% fill) \\
Wall clock        & 31.8\,h \\
Peak memory       & 24.2\,GB \\
Final loss        & 0.114 \\
Cost              & \$86 \\
\bottomrule
\end{tabular}

\end{table}

\paragraph{Reinforcement-learning rig.}
We built and validated a GRPO training rig before concluding that RL was
unnecessary (\S\ref{sec:training}).  The rig used a two-GPU topology: one GPU
trained while the other generated rollouts, with kill-and-resume
checkpointing so that preempted runs lost no more than a single episode.
Exact backpropagation through long episodes was verified against ground truth
before any paid run began.  The reward function was validated separately, by
replaying candidate reward assignments against thousands of previously saved
episodes and confirming agreement before committing GPU spend.  50 clean
steps then ran flat, as reported in \S\ref{sec:training}, and the rig was
shelved in favor of the decoding fix.

% ============================================================================
% M — A Verified Handoff, Before and After the Gate
% ============================================================================
\section{A Verified Handoff, Before and After the Gate}
\label{apx:handoff-example}

Figure~\ref{fig:apx-handoff-example} reproduces one real spontaneous handoff
from the SWE-bench Pro evaluation, shown in the exact post-strip form the
fixer received.  This handoff's reproduction claim was replayed in the task's
own sandbox and genuinely failed, so the claim was forwarded intact rather
than deleted.

\begin{figure}[H]
\centering
\footnotesize
\input{figures/fig-apx-handoff-example}
\caption{\textbf{One real handoff, exactly as the fixer received it.}
A spontaneous handoff from the SWE-bench Pro evaluation, reproduced
verbatim after the verify-then-strip stage.  This reproduction claim was
genuine and forwarded intact; $174$ of $249$ claims were false and
stripped.  Line wrapping in the notes is ours; no content is added or
redacted.}
\label{fig:apx-handoff-example}
\end{figure}

% ============================================================================
% N — Compute and Cost Disclosure
% ============================================================================
% Layout only: start N in a fresh column so its heading, prose and pinned
% table stay together. Without this the heading and prose fall at the foot
% of the previous column and Table XVIII lands alone on the next page.
\newpage
\section{Compute and Cost Disclosure}
\label{apx:spend}

Table~\ref{tab:apx-spend} itemizes the measured spend across the three eras
of this project: evaluation, label collection, and training.
The one-off cost of building the system, comprising the label run
(${\approx}\$370$), the fine-tuning run (${\approx}\$86$), and corpus
note generation (${\approx}\$9$), totals roughly \$465.  Against the
\$0.62 per-task saving over the best solo fixer (\$0.757 API-only
versus \$0.137 all-in), that sunk cost amortizes in roughly 750 tasks.

\begin{table}[H]
\centering
\footnotesize
\caption{\textbf{Measured spend.}
Every figure is a measured value from the run ledgers.  The evaluation
era covers all five arms plus infrastructure; the label-run era covers
fresh per-task outcomes; the training era is the single SFT run.
\modelname's entire evaluation contribution was \$1.13 of GPU time
(${\sim}0.4$ cents per task).  Era totals are not summed.}
\label{tab:apx-spend}
\begin{tabular}{@{}lr@{}}
\toprule
Item & \$ \\
\midrule
\multicolumn{2}{@{}l}{\emph{Evaluation era}} \\
Probe (40 attempts)        & 19.24 \\
\fxgpt solo arm           & 151.64 \\
\fxopus solo arm          & 201.25 \\
\fxkimi solo arm         & 28.23 \\
System arm                 & 86.49 \\
Smoke                      & 1.32 \\
Unrecorded partials        & $\approx$11 \\
GPU (all pods)             & $\approx$5.9 \\
Sandbox compute            & 40--55 \\
\quad era total            & $\approx$545--560 \\
\midrule
\multicolumn{2}{@{}l}{\emph{Label-run era}} \\
Fixer API                  & 340.03 \\
GPU pod                    & $\approx$4 \\
Sandbox compute            & $\approx$25 \\
\quad era total            & $\approx$370 \\
\midrule
\multicolumn{2}{@{}l}{\emph{Training era}} \\
Fine-tuning run            & $\approx$86 \\
Corpus note generation     & $\approx$9 \\
\bottomrule
\end{tabular}

\end{table}

% NOTE: the final \balance was removed. Balancing is incompatible with the
% pinned ([H]) appendix layout (see \nobalance above); \raggedbottom gives
% the intended ragged last page instead.

\end{document}